\documentclass[12pt]{article}
\def\AnswerYes{y}
\def\pdflatex{y}                  
\def\ShowLineNumberVersion{n}     
\def\ShowLabelsVersion{n}         
\def\ShowChangesVersion{n}        
\def\ShowAnnotationsVersion{n}    
\def\ShowFigures{y}               
\def\feynVersion{n}               
\def\MakeArXivLinksActive{y}      
\ifx\pdflatex\AnswerYes
   \usepackage{grffile} 
\fi
\ifx\pdflatex\AnswerYes
    \usepackage{color}
\else
    \usepackage[dvips]{color}
\fi
\ifx\pdflatex\AnswerYes
    \usepackage{thumbpdf}    %
\else
    \usepackage[dvips]{thumbpdf}
\fi
\ifx\pdflatex\AnswerYes
    \usepackage[ 
        final,
        breaklinks=true,
        colorlinks=true,           
        pdfborder={0 0 1},    
        citecolor={blue},linkcolor={blue},urlcolor={red},
        citebordercolor={1 0 0},linkbordercolor={0 0 1},urlbordercolor={1 0 0},
        pdfpagemode=UseOutlines,
        bookmarks=true,bookmarksopenlevel=4
        ]{hyperref}         
\else
    \usepackage[dvips,              
        final,
        breaklinks=true,
        colorlinks=true,      
        pdfborder={0 0 1},    
        citecolor={blue},linkcolor={blue},urlcolor={red},
        citebordercolor={1 0 0},linkbordercolor={0 0 1},urlbordercolor={1 0 0},
        pdfpagemode=UseOutlines,
        bookmarks=true,bookmarksopenlevel=4
         ]{hyperref}         
\fi
\ifx\MakeArXivLinksActive\AnswerYes
   \usepackage{xparse}
   \NewDocumentCommand{\arxiv} %
   {r [: u{ [} u{]]} }{[\href{http://arxiv.org/abs/#2}{arXiv:#2}~[#3]]}
   \NewDocumentCommand{\arxivold} {r[]}{[\href{http://arxiv.org/abs/#1}{#1}]}
   \NewDocumentCommand{\arXiv} %
   {r [: u{ [} u{]]} }{[\href{http://arxiv.org/abs/#2}{arXiv:#2}~[#3]]}
   \NewDocumentCommand{\arXivold} {r[]}{[\href{http://arxiv.org/abs/#1}{#1}]}
\else
   \newcommand{\arxiv}[1][]{[#1]}
   \newcommand{\arxivold}[1][]{[#1]}
   \newcommand{\arXiv}[1][]{[#1]}
   \newcommand{\arXivold}[1][]{[#1]}
\fi
\usepackage{doi} 
\usepackage{orcidlink}
\usepackage[UKenglish]{isodate} 
\usepackage[numbers,sort&compress]{natbib}
\ifx\ShowFigures\AnswerYes
   \usepackage{graphicx}          
\else
   \usepackage[draft]{graphicx}   
   \renewcommand{\includegraphics}[2][]{\fbox{#2}}
\fi
\graphicspath{{figures/}{./}}
\usepackage{multirow}                   
\usepackage{amssymb}
\usepackage{amsmath}
\usepackage{bbm} 
\usepackage{xspace}                    
\xspaceaddexceptions{[]}
\usepackage{dcolumn}                    
\usepackage{pbox}

\usepackage{bm}                        

\usepackage{cancel}	                  

\usepackage{arydshln} 

\usepackage{rotating}

\usepackage{afterpage}          
\usepackage{floatpag}           
                                
\floatpagestyle{empty}          
\usepackage{slashed}
\ifx\feynVersion\AnswerYes
   \usepackage{feynmp-auto} 
\fi

\usepackage{chngcntr} 
\counterwithin*{equation}{section} 
\renewcommand{\theequation}{\thesection.\arabic{equation}}
\renewcommand{\labelenumi}{(\arabic{enumi})}

\ifx\ShowLabelsVersion\AnswerYes
   \usepackage[color]{showkeys} 
   \definecolor{refkey}{gray}{.5}   
   \definecolor{labelkey}{gray}{.5} 
   
\fi
\ifx\ShowAnnotationsVersion\AnswerYes
   \newcommand{\comment}[1]{{\scriptsize\sffamily\bfseries{#1}}}
   \newcommand{\margin}[1]{\marginpar{\scriptsize\sffamily\bfseries{#1}}}
   \newcommand{\drafty}{\textbf{Draft version \today} \hfill}
\else
   \newcommand{\comment}[1]{}
   \newcommand{\margin}[1]{}
   \newcommand{\drafty}{}
\fi
\ifx\ShowLineNumberVersion\AnswerYes
   \usepackage{lineno}
   \linenumbers
\fi
\ifx\ShowChangesVersion\AnswerYes
   \usepackage{ulem}
    
   \newcommand{\delete}[1]{\sout{#1}}            
   \renewcommand{\emph}[1]{\textit{#1}}           
\else

   \newcommand{\sout}[1]{}
   \newcommand{\xout}[1]{}

   \newcommand{\delete}[1]{}
\fi

\newcommand{\eg}{\textit{e.g.}\xspace}
\newcommand{\ie}{\textit{i.e.}\xspace}
\newcommand{\etal}{\textit{et al.}\xspace}

\newcommand{\dis}{\displaystyle}

\newcommand{\hqmmm}{\hspace*{-1.0em}}

\newcommand{\hf}{\hspace*{\fill}}

\newcommand{\e}{\mathrm{e}}
\newcommand{\ii}{\mathrm{i}}
\newcommand{\dd}{\mathrm{d}}

\newcommand{\deint}[2]{\dd^{#1}\;\!\! #2\;}

\newcommand{\vectorwithspace}[1]{\vec{#1}\mkern2mu\vphantom{#1}}

\newcommand{\pv}{\vectorwithspace{p}}

\renewcommand{\Re}{\mathrm{Re}}
\renewcommand{\Im}{\mathrm{Im}}

\newcommand{\N}{\mathrm{N}}
\newcommand{\B}{\mathrm{B}}

\newcommand{\absr}{|r_0|}
\newcommand{\absrm}{|r_0^{-1}|}

\newcommand{\LambdaEfimov}{\ensuremath{\Lambda_\ast}}

\newcommand{\xithr}{\xi_{\mathrm{thr}}}
\newcommand{\xiZB}{\xi_{\mathrm{ZB}}}
\newcommand{\kappatwoZB}[1][]{\kappa_{2\mathrm{ZB}}^{-\:#1}}
\newcommand{\kappathr}{\kappa_{\mathrm{thr}}}
\newcommand{\kappaQU}{\kappa_{3\mathrm{QU}}}

\newcommand{\twoB}{\ensuremath{2\B}\xspace}
\newcommand{\threeB}{\ensuremath{3\B}\xspace}

\newcommand{\jth}{{(j)}}
\newcommand{\multi}{\cdot}

\newcommand{\arccot}{\ensuremath{\mathrm{arccot}}}

\newcommand{\calA}{\mathcal{A}}

\newcommand{\calK}{\mathcal{K}} 
 
\newcommand{\calO}{\mathcal{O}} 
 \newcommand{\calR}{\mathcal{R}}

 \newcommand{\calZ}{\mathcal{Z}}

\newcommand{\mytitle}[1]{\begin{center}\LARGE{\textbf{#1}}\end{center}}
\newcommand{\myauthor}[1]{\textbf{#1}}
\newcommand{\myaddress}[1]{\textit{#1}}
\newcommand{\mypreprint}[1]{\begin{flushright}#1\end{flushright}}

\newcommand{\simge}{\hspace*{0.2em}\raisebox{0.5ex}{$>$}
     \hspace{-0.8em}\raisebox{-0.3em}{$\sim$}\hspace*{0.2em}}
\newcommand{\simle}{\hspace*{0.2em}\raisebox{0.5ex}{$<$}
     \hspace{-0.8em}\raisebox{-0.3em}{$\sim$}\hspace*{0.2em}}

\begin{document}
%

\begin{titlepage}
  \setcounter{page}{0} \mypreprint{
    \drafty
    2nd August 2023 \\
    Revised version 10th August 2023\\
    Re-Revised version 19th October 2023\\
    Text-identical to published version Eur.~Phys.~J.~A \textbf{59} (2023) 289
    \hf 10th December 2023
  }
  
  
\mytitle{Universality of Three Identical Bosons\\ with Large, Negative Effective Range}


\begin{center}
  \myauthor{Harald W.\ Grie\3hammer\orcidlink{0000-0002-9953-6512}$^{a,}$}\footnote{Email: hgrie@gwu.edu}
  
  \emph{and} 
  
  \myauthor{Ubirajara van Kolck\orcidlink{0000-0001-5740-40470000-0001-5740-4047}$^{b,c,}$}\footnote{Email: vankolck@ijclab.in2p3.fr}
  
  \vspace*{0.5cm}
  
  \myaddress{$^a$ Institute for Nuclear Studies, Department of Physics, 
  \\The George Washington University, Washington DC 20052, USA}
  \\[1ex]
  \myaddress{$^b$ Universit\'e Paris-Saclay, CNRS/IN2P3, IJCLab, 91405 Orsay, France}
\\
\myaddress{$^c$ Department of Physics, University of Arizona, Tucson AZ 85721, USA}
\end{center}

\vspace*{0.5cm}

\begin{abstract}
``Resummed-Range Effective Field Theory'' is a consistent nonrelativistic Effective Field Theory of contact interactions with large scattering length $a$ and an effective range 
$r_0$ large in magnitude but negative. Its leading order is non-perturbative, and its observables are universal in the sense that they depend only on the dimensionless ratio $\xi:=2r_0/a$ once the overall distance scale is set by $|r_0|$. 
In the two-body sector, the relative position of the two shallow $S$-wave poles in the complex plane is determined by $\xi$. 
We investigate three identical bosons at leading order for a two-body system with one bound and one virtual state ($\xi\le0$), or with two virtual states ($0\le\xi<1$).
Such conditions might, for example, be found in systems of heavy mesons. 
We find that no three-body interaction is needed to renormalise (and stabilise) the leading order. A well-defined ground state exists for $0.366\ldots\ge\xi\ge-8.72\ldots$. Three-body excitations appear for even smaller ranges of $\xi$ around the ``quasi-unitarity point'' $\xi=0$ ($|r_0|\ll|a|\to\infty$) and obey discrete scaling relations. We explore in detail the ground state and the lowest three excitations. We parametrise their trajectories as function of $\xi$ and of the binding momentum $\kappa_2^-$ of the shallowest \twoB state. These stretch from the point where three- and two-body binding energies are identical to  the point of zero three-body binding. 
As $|r_0|\ll|a|$ becomes perturbative, this version turns into the ``Short-Range EFT'' which needs a stabilising three-body interaction and exhibits Efimov's Discrete Scale Invariance. 
By interpreting that EFT as a low-energy version of Resummed-Range EFT, we match spectra to determine Efimov's scale-breaking parameter $\LambdaEfimov$ in a renormalisation scheme with a ``hard'' cutoff. 
Finally, we compare phase shifts for scattering a boson on the two-boson bound state with that of the equivalent Efimov system.
\end{abstract}


\end{titlepage}

\setcounter{footnote}{0}

\newpage

%

\section{Introduction}
\label{sec:introduction}

The three-boson system displays a remarkable universality when the two-boson subsystem is near the unitarity limit, where the inverse scattering length $a^{-1}$, the effective range $r_0$ and other effective-range-expansion parameters are small on the scale set by the relative momentum. At that point, the two-body ($\twoB$) system is scale invariant, as reflected in a zero-energy pole of the scattering amplitude, and the three-boson system has Discrete Scale Invariance in the form of a geometric tower of so-called Efimov states~\cite{Efimov:1970zz, Efimov:1971zz,  Efimov:1973awb, Efimov:1978pk}. As $|a^{-1}|$ increases, these states display the remarkable twin properties that they are bound (dubbed Borromean binding) even when the $\twoB$ subsystems are unbound ($a<0$), and that they become unstable as attraction increases ($a>0$). The Efimov effect also occurs in multi-component fermions. For example, light nuclei are perturbatively close to unitarity~\cite{Konig:2016utl,Konig:2019xxk}. Efimov states have attracted considerable attention from theory and experiment alike~\cite{Braaten:2004rn,Naidon:2016dpf}.

The Efimov effect takes place in a non-relativistic system where the \twoB scattering length $|a|\gg R$ is much larger than the interaction range $R$, while the other effective-range parameters, and especially $r_0$,  are comparable to $R$. In this article, we consider the universal properties of the three-boson system when the $\twoB$ effective range is large as well, $|r_0|\gg R$. While the Efimov effect appears when the depth of an attractive \twoB potential is precisely adjusted, the situation envisioned here arises when the potential has attractive and repulsive parts that nearly balance each other. The consequence is that the \twoB system has two low-energy $S$-wave poles, whose positions on the complex relative-momentum plane depend on the relative magnitudes and signs of $a$ and $r_0$, encapsulated in the ratio
\begin{equation}
\label{parratio}
   \xi := \frac{2r_0}{a}\;\;.
\end{equation}
This is in contrast to the single (bound or virtual) pole of the near-unitarity case, and this changes the structure of more-body systems dramatically.

While two-body systems with effective ranges which are large in magnitude and negative appear to be an exception, they do occur most likely in scattering systems with a $k\cot\delta$ with large negative curvature (\ie~effective range) and hence typically a resonance just above zero momentum. There are indications of their existence, for example, in the $D^*_s(2317)$ described as bound state of the $DK$-$D_s\eta$ coupled-channel system~\cite{Matuschek:2020gqe}. Some of the $\Xi\N$ and $\Xi\Xi$ two-fermion systems may show large $\xi<0$ as well~\cite{Gasparyan:2011kg, Haidenbauer:2015zqb, Haidenbauer:2021zvr}. It is unclear, however, whether those systems can be treated at sufficient low energies in a single-channel approach where higher effective-range parameters (such as the shape parameter) are not large.

The consequences of shallow poles can be captured by an Effective Field Theory (EFT) designed for relative momenta $k\ll R^{-1}$; see \eg ref.~\cite{Hammer:2019poc} for a recent review. It contains all possible contact interactions, with a finite number of them contributing non-perturbatively at leading order (LO), and the remainder entering perturbatively at higher orders. The existence of a single shallow pole near unitarity requires a single, non-derivative \twoB contact interaction at LO. The three-boson system is not, however, well defined~\cite{Thomas:1935zz}, unless the non-derivative three-body ($\threeB$) contact interaction is also present at LO to ensure consistency with the renormalisation group~\cite{Bedaque:1998kg,Bedaque:1998km}. This \threeB interaction breaks continuous scale invariance down to a discrete subgroup, generating the Efimov effect. The related properties of more-body systems continue to be studied extensively. We will refer to this version as ``Short-Range EFT'', which in Nuclear Physics takes the form of ``Pionless EFT'' and ``Halo/Cluster EFT'', depending on the elementary degrees of freedom being, respectively, just nucleons or including also tight nucleon clusters. In Atomic Physics, it is sometimes also referred to as ``Contact EFT''.

To accommodate a second shallow two-body pole, the two-derivative \twoB contact interaction must be included non-perturbatively at LO~\cite{Habashi:2020qgw,vanKolck:2022lqz}. We call this the ``Resummed-Range EFT''. Renormalisability then constrains~\cite{Phillips:1997xu,Beane:1997pk} the effective range to be negative, 
\begin{equation}
    r_0<0\;\;,
\end{equation}
so that the Wigner bound~\cite{Wigner:1955zz} is satisfied~\cite{Fewster:1994sd,Phillips:1996ae}. Consequently, this EFT admits five combinations of poles of the scattering amplitude on the complex relative-momentum plane:
\renewcommand{\labelenumi}{(\roman{enumi})}
\begin{enumerate}
\item the resonance case ($\xi>1$), with two poles on the lower half-plane which are symmetric with respect to the imaginary axis; 
\item a virtual double-pole ($\xi=1$), namely a double pole on the negative imaginary axis; 
\item two virtual states ($0<\xi<1$), that is, two single poles on the negative imaginary axis; 
\item one virtual and one quasi-bound state ($\xi=0$), \ie~a single pole on the negative imaginary axis and another at the origin (``quasi-unitarity''); and
\item one bound and one virtual state (positive/negative imaginary axis, resp.; $\xi<0$).
\end{enumerate}
In cases (iii) to (v), the poles are symmetric with respect to the point $-\ii/|r_0|$ on the imaginary axis.
Typically, one scans through these pole configurations in this sequence as attraction increases~\cite{Habashi:2020qgw}. They can also be described by replacing~\cite{Habashi:2020ofb,vanKolck:2022lqz} the two-derivative contact interaction by a dimeron field~\cite{Kaplan:1996nv}. The dimeron also accounts more naturally for two poles whose position is nearly symmetrical with respect to the origin, including the case of narrow resonances \cite{Bedaque:2003wa,Gelman:2009be,Alhakami:2017ntb}. That is a particular case of (i) where the poles are much closer to the real axis than to the imaginary one, resulting in a sharp peak in the scattering cross section. The constraint $r_0<0$ follows if one imposes the standard sign for the dimer kinetic term in the Lagrangian, \ie~that the dimer not be a ``ghost''. 

We will describe later that \threeB bound states exist only for cases (iii) to (v), and only for a range of $\xi$ around zero. Of particular interest is case (v), where a third boson can scatter on the two-boson bound state. This is also the situation in Weinberg's discussion of ``elementary'' particles~\cite{Weinberg:1965zz} which is widely employed~\cite{Brambilla:2019esw} in the context of new exotic hadronic states since the discovery of the $X(3872)$~\cite{Belle:2003nnu}. The \twoB sector was analysed from the EFT perspective in ref.~\cite{vanKolck:2022lqz}. We look here for \threeB bound states which emerge as poles of the homogeneous version of the Faddeev equation~\cite{Faddeev:1960su,Eyges:1961}.

We also discuss Borromean systems which emerge for cases (iii) and (iv) ($0\le\xi<1$), namely bound \threeB systems whose \twoB subsystems are unbound into two virtual states. While there is then no scattering on a \twoB state, our equations remain valid but a formal study of \threeB scattering is not presented here. 

In Resummed-Range EFT, the faster fall-off of the \twoB amplitude with momentum suggests that no \threeB interaction is necessary to renormalise the \threeB problem at LO, and indeed our results converge as the cutoff regulator increases. The \threeB spectrum and scattering are thus predicted uniquely, up to corrections from higher-order effective-range parameters. At LO, the magnitude $|r_0|$ of the \twoB effective range fixes the overall scale of the spectrum, and the relative positions of the various levels is universal in the sense that it depends solely on $\xi$. In the ``quasi-unitarity'' limit $|\xi|\ll 1$, we expect an approximately geometric tower of states truncated from below at binding momentum $\kappa_3\sim \absrm$.  Away from this limit, more significant effects of $r_0$ are expected. As $|\xi|$ increases, we will confirm that the spectrum differs from Efimov's, with levels disappearing at different points either at the $\twoB$ threshold (for $\xi<0$) or at vanishing \threeB binding (for $\xi>0$). We will in particular address the question whether the \threeB ground-state energy vanishes before the critical value $\xi=1$ of case (ii) is reached, namely the point at which the two \twoB virtual states coalesce as a precursor to the appearance of a  resonance. We will also see that as either $|\xi|$ or the relative momentum $k$ increases, scattering phase shifts deviate more dramatically from Efimov's.

Frequently, the consequences of Efimov universality are obtained from finite-range potentials whose effective-range effects are made small---see \eg~the highly precise calculations of ref.~\cite{Deltuva:2010xd}. Some of our results could no doubt be extracted from calculations with specific potentials with small shape parameters. However, typically Efimov universality calculations are done with potentials (such as a Gau\3ian, see for example refs. \cite{Gattobigio:2013yda,Recchia:2021snn}) where the \twoB effective range is positive. Here, we isolate instead the effects of a \emph{negative} effective range, imposed by renormalisability of the EFT at $\twoB$ level. This avoids \twoB poles  in the upper complex-momentum half-plane which are difficult to interpret: those not on the imaginary axis~\cite{Hu:1948zz,Schuetzer:1951}, and those on the imaginary axis that are not associated with bound states (``redundant poles'')~\cite{Ma:1946,TerHaar:1946,Ma:1947zz}. Nevertheless it has been suggested~\cite{Beane:2000fi} that the effective ranges in the two-nucleon system, which are positive and somewhat large, should be accounted for at LO through the use of dibaryon fields. Unfortunately, problems then appear in the  three-nucleon system~\cite{Gabbiani:2001yh, Ebert:2021epn}, unless the spurious pole is projected out~\cite{Ebert:2021epn, Ebert:2023aio}, or unless the dibaryon is replaced~\cite{Timoteo:2023dan} by a nonlocal potential~\cite{SanchezSanchez:2020kbx,Peng:2021pvo,Beane:2021dab} of unclear ontological status. In contradistinction, we find no pathologies when $r_0<0$, and its Resummed-Range EFT appears well-defined.

Resummed-Range EFT includes several features of models used for ultracold atoms~\cite{Braaten:2007nq}. In particular, the LO of Resummed-Range EFT describes ``narrow Feshbach resonances'', namely those Feshbach resonances~\cite{vandeKraats:2022kde} that can be modelled as a single channel with large scattering length and negative effective range~\cite{PhysRevLett.93.143201}. Some of our results have already been obtained in this context, and we compare to these as appropriate. There, the dimer field is thought to represent the closed channel~\cite{Gogolin:2008NN}, the quasi-unitarity limit ($|\xi|\ll 1$) is dubbed ``small detuning'' and the opposite limit ($|\xi|\gg 1$) ``intermediate detuning''. We find general agreement with the results reported in ref.~\cite{PhysRevLett.93.143201} for the emergence of the lowest states and in ref.~\cite{Gogolin:2008NN} for excitations, but emphasise the differences with Short-Range EFT. Resummed-Range EFT can account for further interactions at subleading order, starting with the two-body shape parameter one order down in the $kR$ expansion~\cite{Habashi:2020qgw,Habashi:2020ofb}. 

This article is organised as follows. In sect.~\ref{sec:formalism}, we present the formalism for the solution of the Faddeev equation at LO. Section~\ref{sec:results} starts with an overview of our findings for bound states (sect.~\ref{sec:overview}) and then presents results with a discussion of the lowest four bound-state trajectories as function of $\xi$ and of the binding momentum $\kappa_2^-$ of the shallowest \twoB state (sect.~\ref{sec:lowest}), followed by a comparison of these trajectories (sect.~\ref{sec:compare-trajectories}), their parametrisation (sect.~\ref{sec:parametrising}), and finally an interpretation of the Efimov effect in Short-Range EFT as a low-energy version of the Resummed-Range EFT which allows determination of the Efimov parameter (sect.~\ref{sec:compare-Efimov}). Some details on the parametrisation of a state's trajectory are relegated to Appendix~\ref{app:fit}. A presentation of the highlights in scattering (sect.~\ref{sec:scattering}) is followed by the customary summary and outlook in sect.~\ref{sec:conclusions}.

\section{Formalism}
\label{sec:formalism}

We first formulate Resummed-Range (Pionless) EFT at LO in momentum space. We highlight the main features of the \twoB scattering amplitude which are needed as input in the integral equation for the scattering amplitude of a third boson on the \twoB bound state and for the \threeB bound-state equation. We also briefly review the \threeB system in Short-Range EFT and its Efimov physics, for ease of comparison to our results in the next section.

\subsection{The Two-Boson System}
\label{sec:2B}

The LO amplitude of the Resummed-Range EFT arises from the non-perturbative solution of the \twoB system with two contact interactions: one without derivatives, the other with two~\cite{Habashi:2020qgw,vanKolck:2022lqz}. Corrections are systematically accounted for through higher-derivative interactions and included in perturbation theory~\cite{Habashi:2020qgw,vanKolck:2022lqz}. They generate higher effective-range parameters, starting with the shape parameter at next-to-leading order, as well as kinematically relativistic and other effects. This EFT can alternatively be formulated with a dimer field~\cite{Habashi:2020ofb,vanKolck:2022lqz}. We focus on the universal aspects at LO as the input into the \threeB problem.

In this EFT, the LO on-shell amplitude of a system of two identical bosons of mass $M$ and relative momentum $k$ is
\begin{equation}
  \label{eq:2Bamp}
  \calA_{2}(k)=\frac{4\pi}{M}\;
  \frac{1}{\frac{1}{a}-\frac{r_0}{2}\,k^2+\ii \,k}\;\;,
\end{equation}
where $a$ is the scattering length and $r_0\le0$ the effective range. It is convenient to choose one of these two dimensionful parameters to set the overall scale so that systems with the same dimensionless ratio~\eqref{parratio} are self-similar. Since $r_0\le0$ is taken as non-positive throughout, we render momenta  dimensionless by measuring them in units of its magnitude, $\absrm$,
\begin{equation}
  \label{eq:rescaling}
  K:= k\,\absr\;\;.
\end{equation}
This helps track some signs and makes sure that there is no confusion with positive momenta in scattering processes.  The dimensionless version of the above amplitude reads then
\begin{equation}
  \label{eq:2Bamprescaled}
A_{2}(K) := -\frac{M}{4\pi |r_0|} \calA_{2}(k)
=\frac{1}{\frac{\xi-K^2}{2}-\ii K}\;\;.
\end{equation}
The $S$-wave scattering phase shift $\delta(K)$ at momentum $K\ge 0$ is obtained from
\begin{equation}
  \label{eq:2Bamprescaledphaseshift}
 K\cot\delta(K)=A_{2}^{-1}(K)+\ii K \;\;.
\end{equation}

This amplitude has two poles, which for $\xi\le1$ ($r_0\le0\le a$ or $a\le2r_0\le0$) are on the imaginary axis of the complex-$K$ plane at
\begin{equation}
  \label{eq:kappa2}
 K_2^\pm=:\ii \,\kappa_2^\pm\;\mbox{ with }\;\kappa_2^\pm:= - \left[1\pm\sqrt{1-\xi}\right]\;\;.
\end{equation}
Following the definition of ref.~\cite{Habashi:2020qgw}, a real $\kappa_2$ represents thus a pole in $K$ on the imaginary axis, and a positive one represents a bound state.
The deeper pole $\kappa_2^+$ is on the negative imaginary axis, corresponding to a virtual state with a positive residue for $\ii$ times the $S$ matrix. In contrast, the shallow pole $\kappa_2^-$ lies for $\xi<0$ (namely $r_0<0\le a$) on the positive imaginary axis with a positive residue, representing a bound state, or for $\xi>0$ (namely $a\le2r_0<0$) on the negative imaginary axis, representing a virtual state. Unwrapping the rescaling by $\absr$, the shallow state has a binding energy
\begin{equation}
\label{eq:2BBindingEnergy}
    B_2=\frac{(\kappa_2^-)^2}{M\,r_0^2}>0 \;\; ,
\end{equation}
which for $|\xi|\ll 1$ turns into the familiar effective-range expression $B_2(|\xi|\ll 1)\simeq 1/(Ma^2)$. In the ``quasi-unitarity'' limit $\xi\to0$ ($|a|\to\infty$), $\kappa_2^-\to0$  becomes quasi-bound, \ie~it is no longer a pole of the $S$ matrix. We call this ``quasi-unitarity'' since the other pole, $\kappa_2^+\to-2$ still sets a scale in the \twoB system, in contrast to full unitarity which admits no \twoB scale.

To embed the two-body amplitude into the many-body problem, $\calA_2$ of eq.~\eqref{eq:2Bamp} must be Galilei transformed from the $2\B$ centre of mass to that of a many-body system. Denoting the three-momentum and kinetic energy of the \twoB state inside the many-body system by $\pv$ and  $p_0$, respectively, one therefore replaces $k\to\ii\sqrt{-Mp_0+\pv^2/4}$ in  eq.~\eqref{eq:2Bamp}. For scattering on the bound state, the \twoB wave-function renormalisation is 
\begin{equation}
  \label{eq:res2}
  \calZ_{2}:= \frac{M}{4\pi}
\left[\frac{\partial}{\partial p_0} \calA_2^{-1} \left(\ii\sqrt{-Mp_0+\frac{\pv^2}{4}}\right)\right]_{Mp_0=-B_2M}^{-1}
\hqmmm= \frac{2}{M|r_0|}\left(1- \frac{1}{\sqrt{1-\xi}}\right)=\frac{2}{M|r_0|}\;\frac{1}{1+\frac{1}{\kappa_2^-}}\;\;.
\end{equation}
In the limit $\xi\to 0$, one finds $\calZ_{2} \to 2/(Ma)$.

\subsection{The Three-Boson System}
\label{sec:3B}

We consider now the scattering of a boson on the two-boson bound state at total centre-of-mass energy $E$ and relative momentum $k$ at LO in this theory.  The corresponding amplitude is dominated by the $S$ wave. It can sustain \threeB bound states.

The Faddeev equation~\cite{Faddeev:1960su,Eyges:1961} for the $S$-wave amplitude with off-shell momentum $p$ is
  \begin{align}
  \label{eq:3Bamp}
  t(ME;k,p)=&\; \frac{4\pi}{kp}\ln
  \frac{p^2+k^2+pk-ME}{p^2+k^2-pk-ME}\\
  &-\frac{2}{\pi}\int\limits_0^\infty\deint{}{q}\frac{q}{p}\;\ln
  \frac{p^2+q^2+pq-ME}{p^2+q^2-pq-ME}\;
  \frac{t(ME;k,q)}{\frac{1}{a}-\sqrt{\frac{3q^2}{4}-ME}-\frac{r_0}{2}(ME-\frac{3q^2}{4})}\;\;,\nonumber
  \end{align}
where the correct cut structures are made obvious by replacing the real variable $ME$ by $ME+\ii\epsilon$. After rescaling to dimensionless momenta as before and including for the energy 
\begin{equation}
  \label{eq:rescalingE}
  \mu^2:= ME\,r_0^2 \;\;,
\end{equation}
the amplitude at (rescaled) on-shell momentum $K$ and off-shell momentum $Q$ becomes 
  \begin{align}
  T(\mu^2;K,P):=&\;\frac{t(E;k,p)}{r_0^2 }\nonumber
  \\
  \label{eq:3Bamprescaled}
  =&\;\frac{4\pi}{KP}\ln
  \frac{P^2+K^2+PK-\mu^2}{P^2+K^2-PK-\mu^2}
  +\frac{4}{\pi}\int\limits_0^\infty\deint{}{Q}\calK(\mu^2;P,Q) 
  \; T(\mu^2;K,Q)
  \;\;,
  \end{align}
where we introduced the kernel
\begin{equation}
  \label{eq:kernel}
  \calK(\mu^2;P,Q):= \frac{Q}{P}\;\ln
  \frac{P^2+Q^2+PQ-\mu^2}{P^2+Q^2-PQ-\mu^2}\;
  \frac{1}{\xi+(3Q^2-4\mu^2)/4+\sqrt{3Q^2-4\mu^2}}\;\;.
\end{equation}

In scattering ($K>0$) either below or above the three-body threshold, $\mu^2=3K^2/4-(\kappa_2^-)^2+\ii\epsilon$. Accounting for the \twoB wave-function renormalisation of eq.~\eqref{eq:res2}, the relation to the scattering phase shift is
\begin{equation}
  \label{eq:3Bphaseshifts}
  K \cot\delta_3(K)=\;\frac{3\pi}{2} 
  \left[\left(1- \frac{1}{\sqrt{1-\xi}}\right)T(\mu^2;K,K)\right]^{-1}-\ii\,K\;\;.
\end{equation}
In addition, there can be bound-state solutions. Let us denote the dimensionless energy of the $j$th three-body bound state by $\mu^2=-(\kappa^\jth_3)^2<0$, and the corresponding ``binding momentum'' as  $\kappa^\jth_3=\sqrt{|\mu^2|}\in\mathbbm{R}_0^+$ for a bound-state pole at $\ii\,\kappa_3$. By construction and in complete analogy to the \twoB case, this is a  pole which lies on the positive imaginary axis in the complex relative-momentum plane. Bound states are then the nontrivial solutions  of the homogeneous equation with $K=0$,
\begin{equation}
  \label{eq:vertexfu}
  T(\mu^2=-(\kappa^\jth_3)^2;0,P)\;=:\;\Gamma^\jth(P) 
  \propto \int\limits_0^\infty\deint{}{Q}
  \calK(-(\kappa^\jth_3)^2;P,Q)\;\Gamma^\jth(Q)\;\;.
\end{equation}
We choose the proportionality constant for ease of plotting results such that  $\Gamma^\jth(0)=1$. This vertex function characterises the bound-state residue, and hence its wave function. While the two-body system is bound only for $\xi<0$, we will find that \threeB bound states exist also for $\xi=0$, when the shallowest \twoB state is quasi-bound, and for $\xi>0$, when it is virtual. 

\subsection{The Efimov Version: Short-Range EFT}
\label{sec:Efimov}

It is instructive to compare \threeB properties in Resummed-Range EFT with those in the more widely known ``Efimov version'', namely the Short-Range EFT
in which the effective range is considered to be a perturbation~\cite{Hammer:2019poc}. In this EFT, leading order is independent of the effective range, and the \twoB system's amplitude is given by eq.~\eqref{eq:2Bamp} with $r_0=0$ and a single pole at binding momentum $\gamma=1/a$. The  three-boson integral equation is then found by replacing in eq.~\eqref{eq:3Bamprescaled} the  \twoB propagator, 
\begin{equation}
    \label{eq:2BpropEfimov}
    \frac{1}{\xi+(3Q^2-4\mu^2)/4+\sqrt{3Q^2-4\mu^2}}\;\to\;\frac{1}{-2\gamma+\sqrt{3Q^2-4\mu^2}}\;\;,
\end{equation}
where $\gamma\propto 1/a$ is the (rescaled) magnitude of the binding momentum of the \twoB system (pole at $\ii\gamma$). We choose to again rescale each with $\absr$ to render them dimensionless. The resulting \threeB equation is in agreement with, for example, ref.~\cite{Griesshammer:2005ga}.

It is well-explored that for renormalisation this more ultraviolet-sensitive propagator needs a (dimensionless and properly normalised) cutoff-dependent \threeB interaction $H(\Lambda)$ to be added to the  one-body ``exchange potential''~\cite{Bedaque:1998kg,Bedaque:1998km}. The logarithm in  eqs.~\eqref{eq:3Bamprescaled} and~\eqref{eq:kernel} is thus replaced by
\begin{equation}
    \label{eq:2BlnEfimov}
    \ln\frac{P^2+Q^2+PQ-\mu^2}{P^2+Q^2-PQ-\mu^2}\;\to
    \;\ln\frac{P^2+Q^2+PQ-\mu^2}{P^2+Q^2-PQ-\mu^2}
    +\frac{2PQ}{\Lambda^2}\, H(\Lambda)\;\;.
\end{equation}
Like $P,Q$, the cutoff $\Lambda$ is given in units of $\absrm$.
The running of $H(\Lambda)$ with the hard-momentum cutoff $\Lambda$ in the \threeB integral equation is for $\Lambda\gg \gamma$  approximately given by~\cite{Bedaque:1998kg,Bedaque:1998km}
\begin{equation}
\label{eq:Hrunning}
    H(\Lambda)\approx -A\;\frac{\sin[s_0\ln\frac{\Lambda}{\LambdaEfimov}-\arccot s_0]}
    {\sin[s_0\ln\frac{\Lambda}{\LambdaEfimov}+\arccot s_0]}\;\;.
\end{equation}
with $A\approx1$ and $s_0=1.0062378\dots$ the real solution to the transcendental equation
\begin{equation}
\label{eq:s0}
8s_0\;\sinh\frac{\pi s_0}{6}=\sqrt{3}\;\cosh\frac{\pi s_0}{2}\;\;.
\end{equation}
Here, $\LambdaEfimov$ is a scale fixed by some datum, breaking the Continuous Scale Invariance for  $\gamma =0$ down to a discrete one by dimensional transmutation. Like the cutoff $\Lambda$, it is measured in units of $\absrm$. Its precise value depends on the chosen renormalisation scheme, which in our case employs a ``hard'' cutoff regularisation. Its consequence is the Efimov effect~\cite{Efimov:1970zz,Efimov:1971zz,  Efimov:1973awb, Efimov:1978pk}: the \threeB spectrum is geometric, and the binding-momentum ratio between adjacent states is $e^{-\pi/s_0}$. Discrete Scale Invariance is exact at leading order in Short-Range EFT even away from unitarity  for $\gamma \to e^{-\pi/s_0}\gamma$ --- it is broken only if $\gamma$ remains fixed, as in Nuclear Physics. The consequence is that the ``Efimov plot'' of the \threeB binding momentum as function of the \twoB binding momentum $\gamma$ remains self-similar under radial dilations by multiples of $e^{-\pi/s_0}$; see \eg ref.~\cite{Braaten:2004rn}. In particular, the ratio between points where the \threeB states disappear --- either $\mu=0$ for $\gamma<0$ or $\sqrt{|\mu^2|}=\gamma>0$ --- is also $e^{-\pi/s_0}$.

For $\xi\to 0$ in the Resummed-Range EFT we consider here, one \twoB pole becomes very shallow, $\kappa_2^-\to \gamma\to0$, while the other corresponds to momenta comparable to the inverse of the effective range, $\kappa_2^+\to -2$.  For $|\xi|\ll 1$, a r\'egime of momenta 
$|K|\ll |\kappa_2^+|$ exists in which $\kappa_2^+$ only enters as a high-momentum effect. In that case, the \twoB amplitude~\eqref{eq:2Bamprescaled},
\begin{equation}
  \label{eq:2Bamprescaledapprox}
A_{2}(K) = \frac{1}{-\kappa_2^- -\ii K} 
\left[1+ {\cal O} \left(\frac{|K|,\kappa_2^-}{|\kappa_2^+|}\right)\right]\;\;,
\end{equation}
has a single pole $\ii \kappa_2^-$. Thus, in the quasi-unitarity limit $\xi\to 0$ and at sufficiently small momenta, the spectrum and phase shifts should be equally well described by Short-Range EFT 
with a \threeB interaction $H(\Lambda)$ with parameter $\LambdaEfimov$.
Sufficiently far up in the spectrum, the binding momenta  $\kappaQU^{(j)}:= \kappa_3^{(j)} (\xi=0)$ must obey
\begin{equation}
  \label{eq:Efimovratio}
  \lim_{j \to \infty}\frac{\kappaQU^{(j)}}{\kappaQU^{(j+1)}}= 
  e^{\pi/s_0} = 22.6944\ldots\;\;.
\end{equation}

Away from quasi-unitarity, results depend on $\xi$ in principle for all states. Again, these effects should  be small in the evolution of high excitations, for which Short-Range EFT holds. For these, one should thus reproduce consequences of Discrete Scale Invariance even away from unitarity. The values of $\xi$ and $\kappa_2^-$ where the $j$th \threeB state has zero binding, $\xiZB^{(j)}:= \xi(\kappa_{3}^{(j)}=0)$ and $\kappatwoZB[(j)]:=\kappa_2^-(\kappa_{3}^{(j)}=0)$, are expected to obey
\begin{equation}
  \label{eq:Efimovratioprime1}
  \lim_{j \to \infty}\frac{\xiZB^{(j)}}{\xiZB^{(j+1)}}= 
  \lim_{j \to \infty}\frac{\kappatwoZB[(j)]}{\kappatwoZB[(j+1)]}= 
  e^{\pi/s_0} = 22.6944\ldots\;\;.
\end{equation}
Likewise, for the values of $\xi$ and $\kappa_2^-$ where the binding energy of the $j$th \threeB state and the shallow \twoB state coincide, $\xi_\mathrm{thr}^{(j)}:= \xi(\kappa_{3}=\kappa_2^-=:\kappathr)$, 
\begin{equation}
  \label{eq:Efimovratioprime2}
  \lim_{j \to \infty}\frac{\xithr^{(j)}}{\xithr^{(j+1)}}= 
  \lim_{j \to \infty}\frac{\kappathr^{(j)}}{\kappathr^{(j+1)}}= 
  e^{\pi/s_0} = 22.6944\ldots\;\;.
\end{equation}

Since eq.~\eqref{eq:3Bamprescaled} converges for large momenta, we expect in Resummed-Range EFT the existence of a \threeB ground state, in contradistinction to the Efimov effect, where an infinite tower of states of both arbitrarily large and arbitrarily small binding energy exists due to perfect Discrete Scale Invariance. This provides a ``natural'' cutoff to the tower of Efimov states at $\xi=0$. The \twoB effective range $r_0$ (and therefore the existence of the second pole at $\ii\,\kappa_2^+$) also fixes the position of the tower, which in Short-Range EFT is determined by a renormalisation constant $\LambdaEfimov$. Resummed-Range EFT can thus be seen as an underlying theory to Short-Range EFT where one can \emph{predict} $\LambdaEfimov$, rescaled by the remaining dimensionful parameter $r_0$. The deepest bound states should of course show the largest deviation from the Efimov behaviour. As a rough estimation, one might expect departures to scale with $\sim \mathrm{max}(\kappa_2^-,\kappa_3)/|\kappa_2^+|$. Likewise, scattering at momenta $K\simge |\kappa_2^+|$ differs from the one in Short-Range EFT. 

\section{Results for Bound States} 
\label{sec:results}

In this section, we present numerical solutions of the homogeneous equation~\eqref{eq:vertexfu} for bound states. 
Because of the bijection between $\xi$ and $\kappa_2^-$ and because we will compare to the results of Short-Range EFT at the same \twoB binding momentum $\gamma=\kappa_2^-$, we report results  interchangeably as function of $\kappa_2^-=\sqrt{1-\xi}-1$ (eq.~\eqref{eq:kappa2}) and of $\xi=-\kappa_2^-(\kappa_2^-+2)$.

Practically, the bound state is found by discretising the kernel on a quasi-logarithmic mesh in $(P,Q)\in\mathbbm{R}_0^+$ which weighs low momenta exponentially stronger than higher ones, and scanning for nontrivial solutions at particular, discrete $\mu^2=-\kappa_3^2$ of 
\begin{equation}
  \label{eq:det}
  \det[1-\calK(\mu^2;P,Q)]=0 \;\;.
\end{equation}
The vertex function $\Gamma^\jth(Q)$ is then the eigenvector to the kernel with zero eigenvalue. Unless otherwise specified, we use $n=500$ points at a cutoff $\Lambda=45.2548\dots\absrm$. There, residual cutoff-dependence is negligible, and results are quite stable against $n$ or different mappings between the grid and $P,Q$. We consider the four lowest bound states, extrapolate to shallower states, and compare with Discrete Scale Invariance of Short-Range EFT, sect.~\ref{sec:Efimov}. 

\subsection{Bound States: Overview}
\label{sec:overview}

\begin{figure}[t]
\begin{center}
  \includegraphics[width=0.69\linewidth]
  {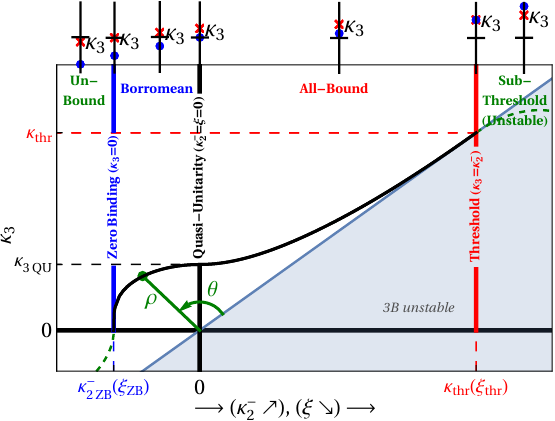}
    \caption{(Colour on-line) Qualitative trajectory of a \threeB bound state as the \twoB binding momentum $\kappa_2^-$ increases --- \ie~$\xi$ decreases --- to the right. Bottom: the four regions and three points of interest; see text. The solid line indicates the \threeB binding momentum, while the green dashed lines in the sub-threshold (unstable) and unbound regions are speculative. In the gray region, the \threeB system is unstable as its binding is smaller than in the \twoB subsystem. The ``polar'' coordinates $(\theta,\rho(\theta))$ of eq.~\eqref{eq:polar} are indicated in green for one point. Top: relative positions of the \twoB pole $\kappa_2^-$ (blue dot) and the \threeB pole $\kappa_3$ (red cross) on the imaginary axis in the complex-$K$ plane.}
  \label{fig:qualitative}
\end{center}
\end{figure}

A qualitative illustration of the trajectories as $\kappa_2^-$ ($\xi$) is  varied is offered in fig.~\ref{fig:qualitative},  where the \threeB binding momentum  $\kappa_3$ is sketched as function of $\kappa_2^-$ (increasing to the right) or $\xi$ (decreasing to the right). The top of the figure shows qualitatively the relative  position of the poles of the \threeB and \twoB bound states in the complex-momentum plane.

The figure leads to a natural classification of the trajectory of each state into four regions, separated by three points of interest whose positions are different for different states:  
\renewcommand{\labelenumi}{(\arabic{enumi})}
\begin{enumerate} 
\item \textbf{Sub-Threshold (Unstable) Region}: For $\kappa_3(\xi)<\kappa_2^-(\xi)$, the \threeB system is less bound than the \twoB one, and thus unstable to the breakup into $1+2$ subsystems. This happens for values $\xi<\xithr<0$ which are negative and sufficiently large in magnitude (\ie~$\kappa_2^->\kappathr>0$ sufficiently large --- the \twoB system in case (v) of the Introduction). 

\item \textbf{Threshold Point}: At some $\xithr<0$, the  \threeB and \twoB systems have identical binding momenta $\kappa_\mathrm{thr}:=\kappa_3(\xithr)\stackrel{!} {=}\kappa_2^-(\xithr)\in\mathbbm{R}^+$ (identical poles on the positive imaginary axis in momentum space), and hence identical binding energies (still case (v) of the Introduction). The \threeB state emerges from the continuum.

\item \textbf{All-Bound Region}: For larger values $0>\xi>\xithr$, $\kappa_3>\kappa_2^->0$, so  that both the \threeB and all \twoB sub-systems are bound (for \twoB: still case (v) of the Introduction).

\item \textbf{Quasi-Unitarity Point}: At $\xi=0$, the binding momentum of the shallowest \twoB bound state vanishes, $\kappa_2^-(\xi=0)=0$, corresponding to an infinite scattering length (case (iv) of the Introduction). This is the point at which comparison with the Short-Range EFT of sect.~\ref{sec:Efimov} is most fruitful. 
The second \twoB state continues to be virtual and has binding momentum $\kappa_2^+(\xi=0)=-2$ in units of $\absrm$, providing a \twoB scale for the \threeB system's binding momentum $\kappaQU:=\kappa_3(\xi=0)\in\mathbbm{R}^+$. 

\item \textbf{Borromean Region}: As $\xi>0$ turns positive, the \threeB system continues to be bound, $\kappa_3(\xiZB>\xi>0)>0$, despite the fact that all \twoB sub-systems become unbound and now contain two virtual states, $\kappa_2^\pm(\xi)<0$ (case (iii) of the Introduction).

\item \textbf{Zero-Binding Point}: The binding momentum (and energy) of the \threeB system becomes zero at some $\xiZB>0$: $\kappa_3(\xiZB)=0$. The \twoB system is already unbound, $\kappatwoZB:=\kappa_2^-(\xiZB)$ with $\kappatwoZB<0$ (case (iii) of the Introduction).

\item \textbf{Unbound Region}: For $\xi>\xiZB>0$ (namely $\kappa_2^-(\xi)<\kappatwoZB<0$), the pole of the \threeB system is no longer on the positive imaginary axis. Both three- and two-body  (cases (i-iii) of the Introduction) systems are unbound.
\end{enumerate}

Finally, we note that 
the \twoB binding momentum has a zero real part, $\Re[\ii\,\kappa_2^\pm]=0$, except in the last region, and only for $\xi>1$. Likewise, the three-body state is bound in all but the last region, and therefore the real part of its pole vanishes, $\Re[K_3^\mathrm{pole}]=\Im[\kappa_3]=0$, while its imaginary part is non-negative, $\Im[K_3^\mathrm{pole}]=\Re[\kappa_3]\ge0$.

The fate of the \threeB trajectory in the sub-threshold and unbound regions will be discussed in future publications~\cite{future}. Close to the threshold, $\xi\simeq  \xi_\mathrm{thr}$, the three-body state is a large ``halo'' system where one boson is on average far from the two-body bound state. Therefore,  the system reduces to a ``core'' whose  constituents are relatively tightly bound, surrounded by another particle which is only loosely bound, and  both interact predominantly in an $S$ wave.  In such an  $S$-wave two-body (core-particle) system without long-range interactions, one expects a single pole in the vicinity of the threshold. When the interaction strength is changed smoothly, continuity dictates that it goes from a bound state (on the positive imaginary axis in momentum) to a virtual state (on the negative imaginary axis in momentum, namely a negative-energy pole on the second Riemann sheet in energy), or vice-versa. This is the case \eg~for an Efimov state~\cite{Rupak:2018gnc}, and we expect the same here. On the other hand, near the zero-binding point $\xiZB$, no \twoB bound state exists, only \twoB virtual states. Like in the sub-threshold region, the \threeB state continues on the imaginary axis as an unbound state. This might be a virtual state, or it might collide with another \threeB pole to acquire a real momentum component, mirroring the behaviour of \twoB poles interpreted as resonance, namely two poles at $\Re[\kappa_3]<0$, $\pm\Im[\kappa_{3}]\ne0$. We cannot exclude the possibility of the collision happening at the origin. In fact, this might be the case in Short-Range EFT, as refs.~\cite{Bringas:2004zz, Deltuva:2020sdd} report a non-vanishing resonance width already just past the zero-binding point. Whatever the exact mechanism, one does not expect a \threeB bound state to be resurrected for $\xi>\xiZB$ or $\xi<\xithr$. Indeed, a wide scan for \threeB bound states in these cases was unsuccessful.

\subsection{The Lowest Three-Body States}
\label{sec:lowest}

\begin{figure}[!t]
\begin{center}
  \includegraphics[width=\linewidth]
  {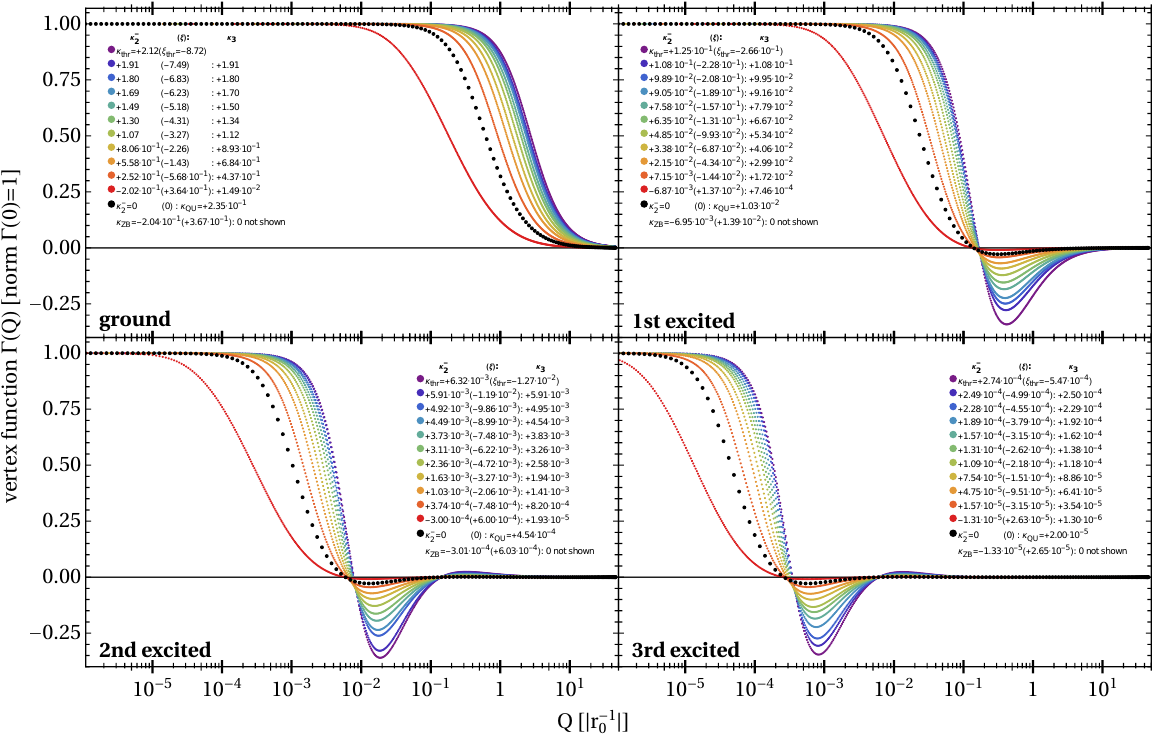}
  \caption{(Colour on-line) Vertex functions of the first four states as functions of momentum in units of $|r_0^{-1}|$, on a logarithmic scale in $Q$. Ten values of $\xi$ are shown in linear spacing between $\xithr$ and $\xiZB$ of the respective state, plus at $\xithr$   and $\xi=0$. The zero-binding state at $\xiZB$ has support only at $Q\to0$ due to its infinite extent in coordinate space and is thus not shown. The legend lists values as $\kappa_2^-\, (\xi)$:$\;\kappa_3$, with binding momenta given in $\absrm$.}
\label{fig:vertexfus-some}
\end{center}
\end{figure}

A lowest-energy (ground) three-body bound state exists only for  $0.367\ldots \ge\xi\ge-8.726\ldots$ (namely $-0.204\ldots\le\kappa_2^-\le2.119\ldots$). Outside this range, a scan for binding momenta $\kappa_3$ in the range $10^2\gtrsim\xi\gtrsim-10^2$ all the way to the cutoff $\Lambda$ and varying $\Lambda$ to several hundred $\absrm$ found no \threeB bound states. This is strong indication that no \threeB bound states exist when the \twoB pole is either a double virtual  pole or a resonance ($\xi\ge1$, $\Re[\ii\,\kappa_2^-]\ne0$). A number of excited bound states do exist, depending on $\xi$.  In this subsection, we present results for the ground state, which we label $j=0$, and the first three excitations, $j=1,2,3$. As to be detailed in sub-sect.~\ref{sec:compare-Efimov}, the trajectory of the $3$rd excitation is near-indistinguishable from the one in Short-Range EFT with the same \threeB binding momentum at quasi-unitarity ($\kappa_2^-=\xi=0$) and can thus be taken to represent the trajectory of an Efimov state.

\begin{figure}[!p]
\begin{center}
  
  \includegraphics[width=\linewidth]
  {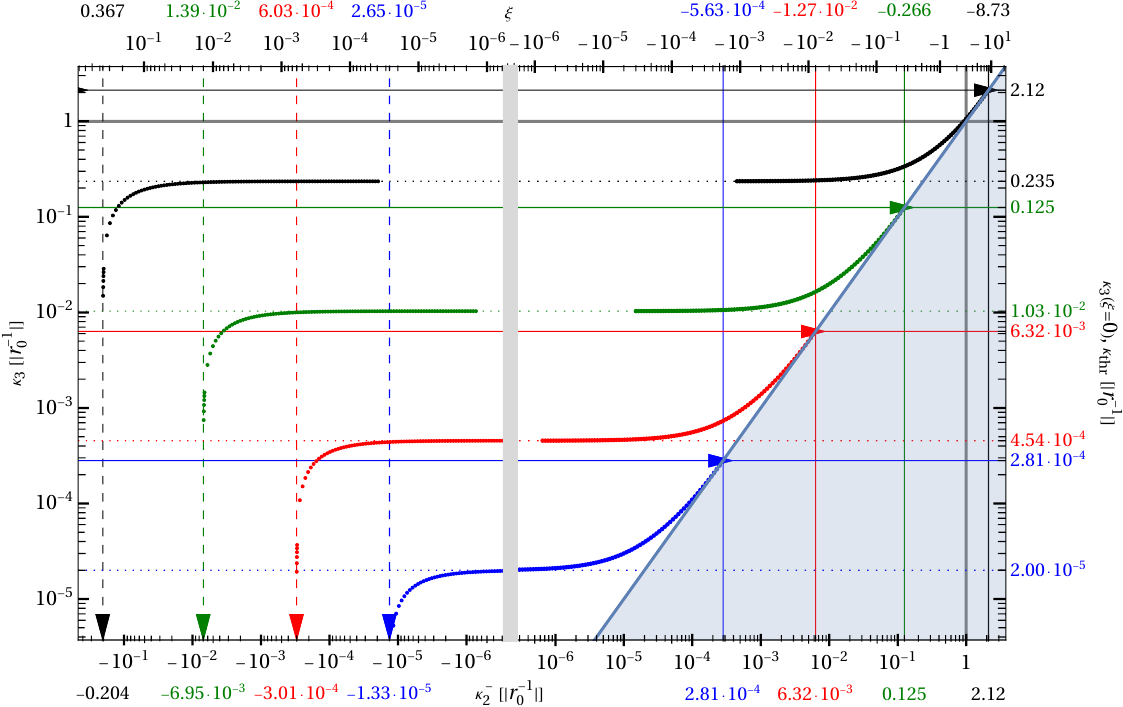}
  
  \caption{(Colour on-line) Binding momenta $\kappa_3$ of the \threeB system in $|r_0^{-1}|$ as function of the \twoB binding momentum $\kappa_2^-$ (bottom scale, in $|r_0^{-1}|$) and of $\xi$ (top scale), for the ground state (black) and the $1$st (green), $2$nd (red) and $3$rd (blue) excitations. A logarithmic scale is used on both axes, with positive and negative values on the $\kappa_2^-$ ($\xi$) axis stitched together by a gray line. In the gray region, a \threeB state is unstable because it has smaller binding energy than the \twoB state, $\kappa_3<\kappa_2^-$, \ie~it disappears into the \twoB continuum. The right axis shows the colour-coded value of the \threeB binding momentum of the corresponding state at the quasi-unitarity limit ($\xi=0$, dotted horizontal lines) and at threshold ($\xithr/\kappathr$, solid horizontal and vertical lines). The top axis also shows the colour-coded values $\xithr$ at threshold and at the zero-binding state, $\xiZB:= \xi(\kappa_3=0)$ (dashed vertical lines); the bottom axis contains the corresponding values of $\kappathr$ and $\kappatwoZB$. Arrows are colour-coded with the corresponding state to mark the points of zero binding ($\kappa_3=0$, vertical arrows); the threshold ($\xithr/\kappathr$, horizontal arrows), and of quasi-unitarity ($\xi=0$, horizontal arrows). Per subsect.~\ref{sec:compare-Efimov}, the trajectory of the $3$rd excitation in this and all subsequent figures is within line thickness indistinguishable from the one in Short-Range EFT with the same \threeB binding momentum at quasi-unitarity ($\kappa_2^-=\xi=0$). }
\label{fig:allbindings}
\end{center}
\end{figure}

\newcommand{\boxing}[2]{\rule[-3.5ex]{0ex}{8.5ex}\pbox{\linewidth}{#1\\[0.4ex]\hspace*{1ex}#2}}
\begin{table}[!p]
  \begin{footnotesize}
 \hspace*{-5ex}%
 \begin{tabular}{|l||l|l|l:l|}
    \hline
    & \multicolumn{1}{c|}{zero binding} & \multicolumn{1}{c|}{quasi-unitarity} & \multicolumn{2}{c|}{threshold}\\
    state& \boxing{$\xiZB:=\xi(\kappa_3=0)$}{($\kappatwoZB:=\kappa_2^-(\xiZB)$)}& $\kappaQU:=\kappa_3(\xi=0)$&$\xithr:=\xi(\kappa_2=\kappa_3)$& $\kappathr:=\kappa_3(\xithr)$\\\hline\hline
    ground \hf($j=0$)        & \boxing{$3.6689(1) \multi 10^{-1}$}{($-2.04318(6)\multi 10^{-1}$)} & $2.35412(3)\multi 10^{-1}$ & $-8.7258(1)$                & $2.11862(2) $ \\\hline
    $1$st excited  \hf($j=1$)& \boxing{$1.3855(1) \multi 10^{-2}$}{($-6.9517(5)\multi 10^{-3}$)} & $1.03030(5)\multi 10^{-2}$ & $-2.6580(1)\multi 10^{-1} $ & $1.25108(1)\multi 10^{-1}$ \\\hline
    $2$nd excited  \hf($j=2$)& \boxing{$6.0279(2) \multi 10^{-4}$}{($-3.0144(1)\multi 10^{-4}$)} & $4.53987(1)\multi 10^{-4}$ & $-1.2680(3)\multi10^{-2} $  & $6.320(1)\multi 10^{-3}$ \\\hline
    $3$rd excited  \hf($j=3$)& \boxing{$2.6538(3) \multi 10^{-5}$}{($-1.3269(2)\multi 10^{-5}$)} & $2.00039(5)\multi 10^{-5}$ & $ -5.632(2)\multi 10^{-4} $ & $2.810(3)\multi 10^{-4}$ \\\hline\hline
    $j\to\infty$ ($\kappa_2^-\to-\frac{\xi}{2}$) & \boxing{\rule[-0ex]{0ex}{4ex}$0.3102(1)\;\e^{-j\frac{\pi}{s_0}}$}{($\rule[-2ex]{0ex}{1ex}\kappa_2^-=-0.1551(1)\;\e^{-j\frac{\pi}{s_0}}$)} & $0.23381(8)\;\e^{-j\frac{\pi}{s_0}}$ & $-6.62(4)\;\e^{-j\frac{\pi}{s_0}}$ & $3.31(2)\;\e^{-j\frac{\pi}{s_0}}$\\\hline
    \hline\hline
    \rule[-1ex]{0ex}{4.8ex}$\frac{\mbox{ground}}{\mbox{$1$st excited}}$ & $26.4808(2) $ & $22.849(1) $ & $32.828(2) $ & $ 16.934(2)$\\
    \hf(rel.~dev.~to $\e^{\frac{\pi}{s_0}})$ & \hf($1.668(1)\multi 10^{-1} $) & \hf($ 6.81(3)\multi 10^{-3}$) & \hf($ 4.4653(1)\multi 10^{-1}$) & \hf($-2.537(1)\multi 10^{-1}$)\\\hline
    \rule[-1ex]{0ex}{4.5ex}$\frac{\mbox{$1$st excited}}{\mbox{$2$nd excited}}$ & $22.9847(2) $  & $22.695(1) $  & $20.963(5) $  & $ 19.795(3)$ \\
    \hf(rel.~dev.~to $\e^{\frac{\pi}{s_0}})$ & \hf($1.279(1) \multi 10^{-2}$) & \hf($ 2.7(4)\multi 10^{-5}$) & \hf($ -7.6529(2)\multi 10^{-2}$) & \hf($ -1.277(2)\multi 10^{-1}$)\\\hline
    \rule[-1ex]{0ex}{4.5ex}$\frac{\mbox{$2$nd excited}}{\mbox{$3$rd excited}}$ & $22.7142(6) $  & $ 22.694(5)$  & $ 22.556(3)$  & $ 22.488(5)$ \\
    \hf(rel.~dev.~to $\e^{\frac{\pi}{s_0}})$ & \hf($8.65(5)\multi 10^{-4}$) & \hf($ 0(4)\multi 10^{-6}$) & \hf($ -7.9(5)\multi 10^{-3}$) & \hf($ -9.0(1)\multi 10^{-3}$)\\\hline\hline
    \rule[-2ex]{0ex}{5.5ex}ratio $j\to\infty:\;\e^\frac{\pi}{s_0}$       & $ 22.6944\dots$  & $22.6944\dots $  & $22.6944\dots $  & $22.6944\dots $ \\\hline
  \end{tabular}
  \end{footnotesize}
  \caption{Characteristics of the point of zero binding energy $\xiZB:=\xi(\kappa_3=0)$, of the quasi-unitarity limit ($\kappa_2^-=\xi=0$) with   binding momentum $\kappaQU:=\kappa_3(\xi=0)$, and of the  threshold point $\kappathr(\xithr):=\kappa_3(\xithr)\stackrel{!}{=}\kappa_2^-(\xithr)$ where the binding energies of  the \twoB and \threeB system are identical. Values are given for the ground state and first three excitations, plus extrapolated values for the Efimov-like states. The lower part lists the ratios of the properties between adjacent states and in parenthesis their relative deviations $\dis\e^{-\pi/s_0}\;\frac{\mbox{\footnotesize property of $(j-1)$st state}}{\mbox{\footnotesize 
  property of $j$th state}}-1$ from the scaling factor $\e^{\pi/s_0}\approx22.6944\ldots$. 
  The imaginary parts $\kappa_3$ of binding momenta are quoted in $\absrm$, and their real parts are zero. Uncertainties combine extrapolation and numerics.}
  \label{tab:characteristics}
\end{table}

The vertex functions $\Gamma^\jth(Q)$ shown in fig.~\ref{fig:vertexfus-some} for a number of  $\xi$ values display the expected behaviour: with decreasing binding energy, the typical momentum is shifted to lower $Q$, making the state wider; and the vertex function of the $j$th excitation has $j$ nodes, just as the operator $1-\calK(\mu^2;P,Q)$ of eq.~\eqref{eq:kernel} also has $j$ negative eigenvalues. 

The respective binding momenta are given in fig.~\ref{fig:allbindings} as function of $\xi$ or $\kappa_2^-$, aligning with the qualitative behaviour of fig.~\ref{fig:qualitative}. The number of bound \threeB states increases as $|\xi|\to0$: none above the \twoB state at $\xi=-10$, but one at $\xi=-1$, two at $\xi=-0.1$, three at $\xi=-0.01$, and four at $\xi=-0.0001$. 
Likewise, states disappear one by one as $\xi$ increases from zero.

Table~\ref{tab:characteristics} lists the values at the points of interest for the first four states, and their ratios between different states. Given are the two thresholds $\xiZB$ and $\xithr$ with their respective  \twoB binding momenta $\kappatwoZB$ and $\kappathr$, as well as the \threeB binding momentum at quasi-unitarity, $\kappaQU$. These results are inferred from computations at $13$ cutoffs, equidistant in $\ln\Lambda$ between $16\,\absrm$ and  $128\,\absrm$, with grids of $500$, $1000$ and $1500$ points, and independent variations of the mappings between the grid  and the momentum variables $P,Q$. The uncertainties are from the $n\to\infty$ and $\Lambda\to\infty$ extrapolation and  numerics combined. Determining the ratios both independently and using the extracted values of $\xi$ and $\kappa_3$ leads to central  values and errors which are consistent with each other. The reported values and uncertainties of all quantities were  determined to be consistent using the network of individual results and from stability considerations. The $\xithr$ values for the ground state and first excitation are consistent with those found in ref.~\cite{PhysRevLett.93.143201}, namely $\approx 8.8$ and $\approx 0.28$, respectively.

\subsection{Comparison between Trajectories}
\label{sec:compare-trajectories}

We first focus on comparisons between the trajectories of the ground state and first three excitations; subsect.~\ref{sec:compare-Efimov} will compare them to the results of Short-Range EFT. Perhaps the most striking feature of table~\ref{tab:characteristics} and fig.~\ref{fig:allbindings} is the apparent  self-similarity of the different trajectories $\kappa_3^\jth(\xi)$, indicating the approximate invariance under a discrete scaling transformation. 

\begin{figure}[!p]
\begin{center}
  \includegraphics[width=0.688\linewidth,bb = 0 16 319 207,clip=]
  {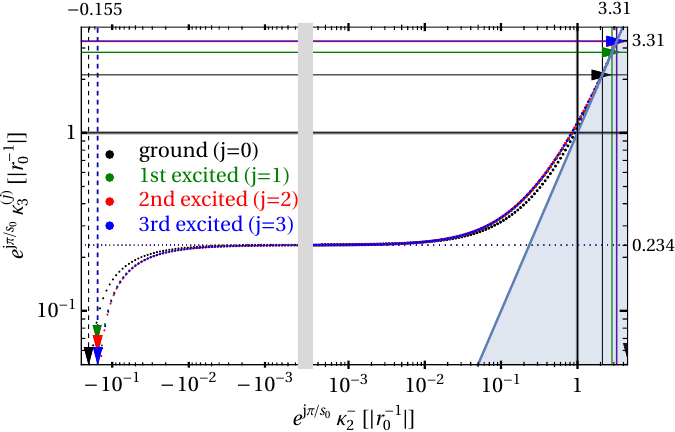}\\
  \hspace*{-5.3ex}\includegraphics[width=0.675\linewidth]
  {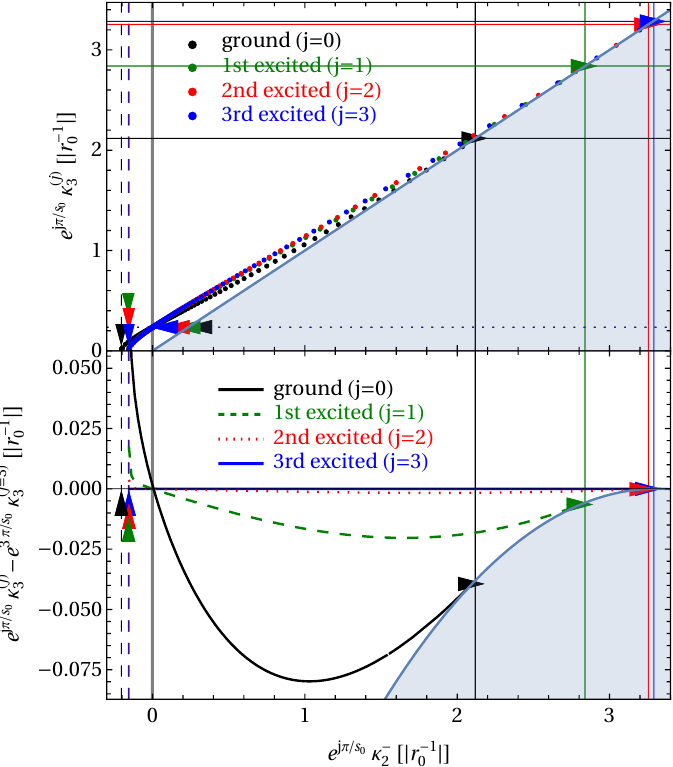}
  \caption{(Colour on-line) The rescaled trajectories of the binding momentum $\e^{j\pi/s_0}\,\kappa_3$ of the $j$th \threeB state as function of the binding momentum $\e^{j\pi/s_0}\,\kappa_2^-$ of the shallow \twoB state, both in units of $|r_0^{-1}|$.
  Top: \emph{logarithmic} scales on both axes, with $\kappa_2^->0$ and $<0$ stitched together, and the top and right axes indicating the universal values at threshold, quasi-unitarity and zero binding for $j\to\infty$ as provided in table~\ref{tab:characteristics}. 
  Centre: \emph{linear} scales on both axes, rendering the different threshold and zero-binding points of the states more prominent.
  Bottom: difference between the $j$th and $3$rd state. 
  Colour-coding, vertical/horizontal lines, arrows and shaded area as in fig.~\ref{fig:allbindings}.
  }
\label{fig:rescaled}
\label{fig:binding-thr}
\end{center}
\end{figure}

For a quantitative comparison, we first rescale the trajectory of the $j$th state, $(\kappa_2^-, \kappa_3)\;\to\;\e^{j\pi/s_0}(\kappa_2^-, \kappa_3)$, by the Efimov ratio expected for $j\to\infty$ from the discussion in sect.~\ref{sec:Efimov}. The result is shown on the top of fig.~\ref{fig:rescaled} on logarithmic scales, and at its bottom using linear scales. The ratios of the points at threshold, quasi-unitarity and zero binding are listed in  table~\ref{tab:characteristics}, together with uncertainties. There are indeed a few noticeable differences: The ground state has the ``shortest'' rescaled trajectory, and length grows with each excitation. While the rescaled zero-binding  point and  \threeB binding momentum at quasi-unitarity remain largely in place as $j$ is increased, subsequent excitations stretch the threshold position by more than half of the ground-state value. The logarithmic scale on the top of fig.~\ref{fig:rescaled} makes one intuitively underestimate this significant effect, which is better captured on the linear scale at the bottom. Asymptotically for $j\to\infty$, the trajectory of $\kappa_3^{-\,\jth}$ stretches between zero binding at $\kappatwoZB[(j\to\infty)]=-0.155\dots\;\e^{-j\pi/s_0}$ and a threshold at $\kappathr^{(j\to\infty)}=3.3\dots\;\e^{-j\pi/s_0}$, or minus-twice those values in $\xi$ since $\kappa_2^-(\xi)\to-\xi/2$ for $|\xi|\to0$. Similarly, quasi-unitarity is reached for $\kappaQU^{(j\to\infty)}=0.2338\dots\;\e^{-j\pi/s_0}$.

The main differences among trajectories are found where the state goes unbound or unstable, and for the ground state. To highlight them, we re-plot in fig.~\ref{fig:binding-thr} the same data on linear scales as differences $\e^{j\pi/s_0}\,\kappa_3^\jth-e^{3\pi/s_0}\,\kappa_3^{(0)}$ between the rescaled binding momentum of the $j$th state and  the shallowest state investigated, $j=3$, as function of $\e^{j\pi/s_0}\,\kappa_2^-$. This plot shows again that the differences are  large towards the threshold points, and non-negligible towards zero binding. However, little difference occurs around quasi-unitarity, where the three curves very nearly intersect at one point for $\kappa_2^-=\xi=0$. Deviations are most striking for the ground state,  visible for the first excitation, and barely noticeable for the second one. Remarkably, though, they are not very large even for the ground state. Note that for $\kappa_2^-\simeq\kappatwoZB$ ($\xi\simeq \xiZB$), differences are artificially enhanced by the different points where each state goes unbound, since $e^{3\pi/s_0}\,\kappa_3^{(0)}\to0$ before $\e^{j\pi/s_0}\,\kappa_3^\jth$ does.

\begin{figure}[!t]
\begin{center}
  \includegraphics[width=\linewidth]
  {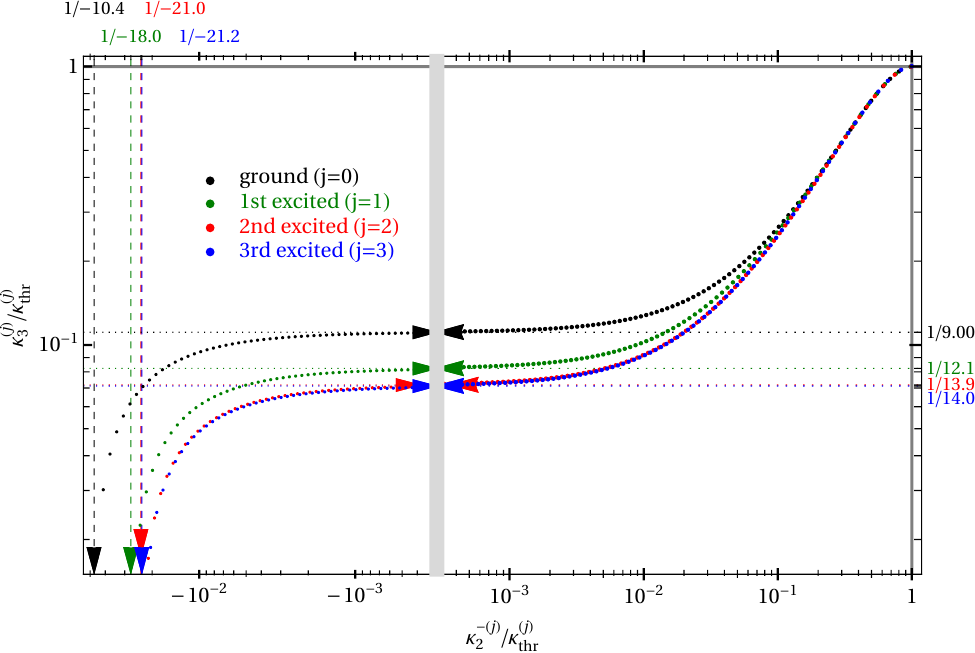}
  \caption{(Colour on-line) The universal function  $\calR^\jth(x_\kappa):=\kappa_3^\jth/\kappathr^\jth$ of the binding momentum of the $j$th state as function of the rescaling variable $x_\kappa:=\kappa_2^-/\kappathr^\jth$ of  the respective state, both on a logarithmic scale. The inverse of the ratio for $\kappaQU$ at quasi-unitarity ($\kappa_2^-=\xi=0$) and for $\kappatwoZB$ at   zero binding ($\kappa_3=0$) are given on the top and left, as provided   in the second and last column of table~\ref{tab:characteristicsTrajectory}, respectively. Colour-coding, vertical/horizontal lines, arrows and $\kappa_2^->0$ and $<0$ stitched  together as in fig.~\ref{fig:allbindings}. 
  }
\label{fig:trajectories-kappa3-kappa2}
\end{center}
\end{figure}

Another way to characterise differences and commonalities of trajectories is to rescale the \twoB and \threeB binding momenta by their respective value $\kappathr^\jth$ at threshold, defining 
\begin{equation}
  \label{eq:kappa2-ratio}
  \calR^\jth\left(x_\kappa:=\frac{\kappa_2^-}{\kappathr^\jth}\right):=
  \frac{\kappa_3^\jth(\kappa_2^-)}{\kappathr^\jth}\;\;.
\end{equation}
These rescaled trajectories are shown in fig.~\ref{fig:trajectories-kappa3-kappa2} as function of $x_\kappa$. Their characteristic values at zero binding and  quasi-unitarity are given in table~\ref{tab:characteristicsTrajectory} for the first four states within the same trajectory and extrapolated for $j\to\infty$, derived as described for table~\ref{tab:characteristics}.
The function $\calR^\jth$ is not only universal for any given state, but it also appears to approach a common, universal function for shallower states. While differences are clearly seen for the ground state and first excitation, already for the second excitation values are very close to those for the third one. A similar plot results if the rescaled trajectories are displayed as function of the rescaled variable $x_\xi:= \xi/\xithr$. However, some differences arise from the nontrivial mapping between $\xi$ and $\kappa_2^-$, which only approaches a simple rescaling $x_\kappa\to x_\xi$ from $\xi\to-2\kappa_2^-$ as $\xi\to0$. 

\subsection{Parametrising Trajectories}
\label{sec:parametrising}

These figures in combination highlight two striking features which are confirmed by a closer numerical analysis. First, the  trajectories as function of $x_\kappa$ are extremely similar towards threshold, $x_\kappa=1$. A fit
\begin{equation}
    \label{eq:trajectoryToThreshold}
    \calR(x_\kappa\to1)=x_\kappa+0.098(1)\;(1-x_\kappa)^2+\calO((1-x_\kappa)^3)
\end{equation}
is right between the $j=0,1$ trajectories which are bit higher, and the $j=2,3$ trajectories. The error includes the uncertainties of the  individual fit and is stable against including more fit terms. The relative residuals  $\frac{\mathrm{data}(x_\kappa)-\mathrm{param}(x_\kappa)}{\mathrm{data}(x_\kappa)}$  increases from $0$ by construction at threshold to $\pm0.4\%$ at $x_\kappa=0.5$ and $\pm1.5\%$ at $x_\kappa=0.3$. This nicely quantifies  the extremely slow convergence between the binding momenta of the \threeB and \twoB systems towards threshold.

The slopes around quasi-unitarity are also quite similar. Fitting
\begin{equation}
    \label{eq:trajectoryToQU}
    \calR^\jth(x_\kappa\to0)=\frac{\kappaQU^\jth}{\kappathr^\jth}+a^\jth\;x_\kappa+\calO(x_\kappa^2)
\end{equation}
with the constants at quasi-unitarity from table~\ref{tab:characteristicsTrajectory} gives noticeable deviations from the straight line  only for $|x_\kappa|>0.005$. The slope increases slightly from the ground state ($a^{(0)}=0.8289(3)$) via the first excitation  ($a^{(1)}=1.035(1)$) to $a^{(j=2\to\infty)}=1.057(1)$. These fits are again quite stable against varying the data range or including higher-order terms. 

The approach to zero binding appears quite uniform for all states, too. An expansion
\begin{equation}
    \label{eq:trajectoryToZB}
    \calR^\jth\left(x_\kappa\to\frac{\kappatwoZB[\jth]}{\kappathr^\jth}\right)=\left(x_\kappa-\frac{\kappatwoZB[\jth]}{\kappathr^\jth}\right)^{\alpha\jth}
\end{equation}
leads to approximately the same exponent $\alpha(j)\approx0.58(3)$ for each state, but numerical stability is more elusive. The exponent grows as the fit range is increased, while the residuals remain comparable.

Finally, we parametrise the whole trajectory $\calR^\jth(x_\kappa)$ of each state, not just the behaviour at the three  ``critical'' points. We are inspired by Efimov's universal function $\Delta(\xi)$~\cite{Efimov:1971zz,  Efimov:1973awb, Efimov:1978pk} to transform the curve from Cartesian coordinates in the $(x_\kappa, \calR^\jth(x_\kappa))$ plane to polar ones,
\begin{equation}
    \label{eq:polar}
    \begin{split}
    x_\kappa&=\sqrt{2}\;\rho^\jth(\theta)\;\cos(\theta+\frac{\pi}{4})\;\;,\\
    \calR^\jth(x_\kappa)&=\sqrt{2}\;\rho^\jth(\theta)\;\sin(\theta+\frac{\pi}{4})
    \;\;,
    \end{split}
\end{equation}
as shown in fig.~\ref{fig:qualitative}. 
We choose $\theta$ to measure the angle formed by the threshold point, the origin, and any particular point on the trajectory (chosen in the mathematically positive sense and continuous between the first and second quadrant),
\begin{equation}
    \label{eq:polartheta}
    \theta:=\arctan\frac{\calR^\jth(x_\kappa)}{x_\kappa}\;-\;\frac{\pi}{4}\;\;.
\end{equation}
By our choice of normalisation and using $\calR(x_\kappa\to1)\to 1$, the radius defined as 
\begin{equation}
    \label{eq:polarrho}
    \rho^\jth(\theta):=\frac{1}{\sqrt{2}}\;\sqrt{x_\kappa^2(\theta)+(\calR^\jth(\theta))^2}
\end{equation}
is rescaled to $1$ at threshold.
The coordinates of the points of interest and regions are then
\begin{equation}
    \begin{array}{lll}
    \mbox{Threshold Point:}&\dis\theta_\mathrm{thr}=0&\mbox{ with 
    }\dis\rho^\jth_\mathrm{thr}=1\\
    \mbox{All-Bound Region:}&\dis\theta\in]0;\frac{\pi}{4}[&\\
    \mbox{Quasi-Unitarity Point:}&
    \dis\theta_\mathrm{QU}=\frac{\pi}{4}&\mbox{ with }\dis\rho^\jth_\mathrm{QU}=\frac{\kappaQU^\jth}{\sqrt{2}\,\kappathr^\jth}\\
    \mbox{Borromean Region:}&\dis\theta\in]\frac{\pi}{4};\frac{3\pi}{4}[&\\[1ex]
    \mbox{Zero-Binding Point:}&
    \dis\theta_\mathrm{ZB}=\frac{3\pi}{4}&\mbox{ with }\dis\rho^\jth_\mathrm{ZB}=\left|\frac{\kappatwoZB[\jth]}{\sqrt{2}\,\kappathr^\jth}\right|
    \end{array}
\end{equation}

Concretely, we are inspired by an idea for $\Delta(\xi)$ by Gattobigio \etal~\cite{Gattobigio:2019eqw} to parametrise
\begin{equation}
\label{eq:parametrisation}
    \rho^\jth_\mathrm{param}(\theta)=\exp\left(\sum\limits_{n=1}^7\;c_n^\jth\;\theta^{n/2}\right)\;\;.
\end{equation}
The powers of square-roots are essential to capture the behaviour towards the threshold, $\theta\to0$. Gattobigio \etal~use a slightly different definition of the angle and  have one more parameter since the scale of their function is not fixed at one point. Remember that we chose $\rho^\jth(0)=1$ for all states by normalisation, and the scale can be restored by multiplying with the threshold value,  $\kappathr^\jth$. We find a more economical parametrisation with an exponential function of powers. Without an exhaustive search, we did not find more efficient combinations of powers of roots, but a (non-exponentiated) $7$th degree  polynomial in $x^{n/4}$ and with 
$c_0^\jth=1$ comes close. We also employ a variant of the idea by Gattobigio \etal~to construct a nearly-best-fit whose coefficients have exactly a given number of significant figures; details on the procedure and its quality can be found in Appendix~\ref{app:fit}.

\begin{table}[!b]
  \centering
  \begin{footnotesize}
    \begin{tabular}{|r||c|c|c|c|c|c|c||c|}
    \hline
    state&$c_1^\jth$&$c_2^\jth$&$c_3^\jth$&$c_4^\jth$&$c_5^\jth$&$c_6^\jth$&$c_7^\jth$&$w_\mathrm{pred}^\jth$\\\hline\hline
    ground ($j=0$)       &$-4.7545$&$ 4.3710$&$ -7.4404$&$ 11.714$&$ -11.0015$&$ 5.54726$&$-1.12595$   &$0.00009$\\\hline
    $1$st excited ($j=1)$&$-4.2764$&$ 1.0734$&$ -1.6005$&$ 5.4777$&$ -6.58104$&$ 3.64577$&$-0.768214$  &$0.00003$\\\hline
    $2$nd excited ($j=2)$&$-4.6823$&$ 1.4352$&$ -1.2221$&$ 4.1199$&$ -5.08001$&$ 2.87318$&$-0.613720$  &$0.00003$\\\hline
    $3$rd excited ($j=3)$&$-4.7003$&$ 1.1952$&$ 0.014514$&$ 1.5657$&$-2.44535$&$ 1.53708$&$-0.348975$  &$0.00011$\\\hline
    \end{tabular}
    \caption{The coefficients of the parametrisation $\rho_\mathrm{param}^\jth$ of the $j$th state in eq.~\eqref{eq:parametrisation}, in $\mathrm{rad}^{-n/2}$, with $5$ and $6$ significant figures, respectively; see text and Appendix~\ref{app:fit}. In the last column, $w^\jth_\mathrm{pred}(\theta)$ estimates the $68\%$ confidence interval $\rho_\mathrm{param}^\jth(\theta)\pm   w_\mathrm{pred}^\jth$ for a prediction at a single, new $\theta$.} 
    \label{tab:parametrisation}
    \end{footnotesize}
\end{table}

Table~\ref{tab:parametrisation} contains the parameter values for the ground state and first three excitations from a least-square fit to $208$ points per trajectory, including higher densities of values exactly at and very close to threshold, quasi-unitarity and zero binding. We do not quote parameter uncertainties since all entries of the correlation matrix are bigger than $0.66$ in magnitude, indicating that they are highly (anti-)correlated. A parametrisation with fewer coefficients is therefore likely to exist. The last column in the table contains the half-width of the parametrisation's $68\%$ confidence prediction interval $\rho_\mathrm{param}^\jth(\theta)\pm w_\mathrm{pred}^\jth$ for the likely position of a new datum at one single $\theta$. This includes the variation on both the parameters and the data scatter (variation of ``future'' individual value).  Within uncertainties, it can be treated as angle-independent. The half-width of the $95\%$ interval is well-approximated by $2w_\mathrm{pred}^\jth$.

\begin{figure}[!b]
\begin{center}
  \includegraphics[width=0.7\linewidth]
  {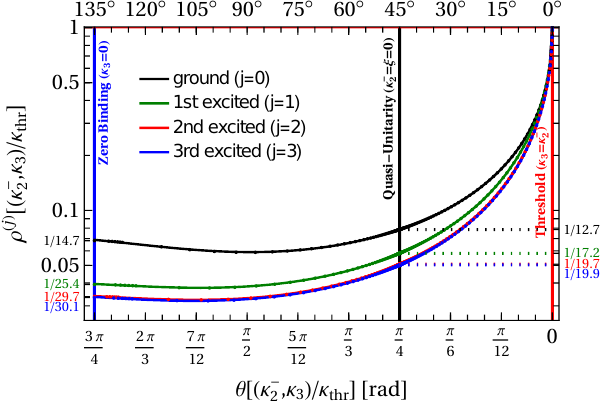}
\caption{(Colour on-line) The data and fitted function of trajectories of the ground state and first three excitations in polar form,  eq.~\eqref{eq:polar}, in a linear-log plot. For ease of comparison with the other figures, the angle $\theta$ increases to the left, \ie~the threshold is on the right. Therefore, slopes have the opposite sign to what the figure seems to imply at first glance. 
Values at zero binding are indicated on the left, those at quasi-unitarity on  the right, each in the colour of the corresponding state. As detailed in subsect.~\ref{sec:compare-Efimov}, the blue solid line of the $3$rd excitation is indistinguishable from the result in Short-Range EFT with the same \threeB binding momentum at quasi-unitarity ($\kappa_2^-=\xi=0$).}
\label{fig:parametrisation}
\end{center}
\end{figure}

Figure~\ref{fig:parametrisation} shows the result, together with the values of $\rho^\jth$ at points of interest. A detailed discussion of the fit accuracy is provided in Appendix~\ref{app:fit}.

The fit coefficients of table~\ref{tab:parametrisation} approach roughly universal values with increasing excitation and appear all of similar, ``natural''  size. This indicates at best limited fine-tuning, despite a rather strong correlation matrix which masks a common limit for the coefficients as $j\to\infty$. 
\begin{table}[!t]
  \centering
  \begin{footnotesize}
    \begin{tabular}{|l@{\hspace*{-16ex}}r||c|c|c|c|c|}
    \hline
    point&property&ground ($j=0$)&$1$st excited&$2$nd excited&$3$rd excited&$j\to\infty$\\\hline\hline
    \underline{threshold:}& $\rho^\jth(0)$ &\multicolumn{5}{c|}{$1$ (normalisation)}\\\hline
    &\pbox{\linewidth}{slope $(\rho^\jth_\mathrm{param)})^\prime(\theta\to0)$\\[-0.5ex]\hspace*{\fill}$[\mathrm{rad}^{-1}]$}&
    \multicolumn{5}{c|}{\rule[-3ex]{0ex}{7.5ex}$\dis\frac{c_1^\jth}{2\;\sqrt{\theta}}\to\infty$}\\
    \hline\hline
    \underline{quasi-unitarity:}& $\rho^\jth_\mathrm{param}(\frac{\pi}{4})$ &$0.078571(7)$&$ 0.058244(2)$&$ 0.050779(2)$&$ 0.050143(8)$&\multirow{2}{*}{$0.0501(2)$}\\
    &from table~\ref{tab:characteristics}&$0.078571(1)$&$ 0.058232(3)$&$ 0.050794(8)$&$ 0.05034(5)$&\\
    &relative residual &$-0.00(1)\%$& $-0.02(1)\%$& $0.03(2)\%$& $0.4(1)\%$&\\
    \hline
    &\rule[-1.5ex]{0ex}{4.5ex}slope $(\rho^\jth_\mathrm{param})^\prime(\frac{\pi}{4})$ $[\mathrm{rad}^{-1}]$ 
        &$ -0.06557(7) $&$ -0.06042(1)$&$ -0.053653(1)$&$ -0.052971(3)$&\multirow{2}{*}{$-0.0530(2)$}\\
        &from $x_\kappa^\jth\to0$ fit & $-0.0651(1)$& $-0.0603(1)$& $-0.0537(1)$& $-0.0532(1)$&\\
        &relative residual & $-0.7(2)\%$& $-0.2(2)\%$& $-0.1(2)\%$& $-0.4(2)\%$&\\
    \hline\hline
    \underline{zero binding:}& $\rho^\jth_\mathrm{param}(\frac{3\pi}{4})$   &$0.06895(5)$&$ 0.03953(2)$&$ 0.03393(1)$&$ 0.03359(7)$&\multirow{2}{*}{$0.0331(2)$}\\
    &from table~\ref{tab:characteristics}&$0.068193(2)$&$ 0.039290(3)$&$ 0.033726(5)$&$ 0.03339(4)$&\\
    &relative residual& $-1.1(1)\%$& $-0.62(5)\%$& $-0.62(5)\%$& $-0.60(2)\%$&\\\hline
    &\rule[-1.5ex]{0ex}{4.5ex}slope $(\rho^\jth_\mathrm{param)})^\prime(\frac{3\pi}{4})$ $[\mathrm{rad}^{-1}]$ 
        &$ 0.0155(4)$&$ 0.00439(2)$&$ 0.00427(2)$&$ 0.0052(5)$&$0.0050(5)$\\
    \hline
    \end{tabular}
    \caption{Values from the parametrisation $\rho_\mathrm{param}^\jth$, eq.~\eqref{eq:parametrisation} and of derivatives at points of interest, compared to those derived from table~\ref{tab:characteristics}. The relative residual is $1-\rho_\mathrm{param}^\jth/\rho_\mathrm{table}$. The extrapolations in the Efimov limit $j\to\infty$ are withing respective uncertainties consistent with the results of tables~\ref{tab:characteristics} and~\ref{tab:characteristicsTrajectory}.} 
    \label{tab:characteristicsPolar}
    \end{footnotesize}
\end{table}

Table~\ref{tab:characteristicsPolar} contains the values of  $\rho^\jth$ at the points of interest in the parametrisation, as well as its values for the derivatives at quasi-unitarity and zero binding. As the table shows, these match the values derived via eq.~\eqref{eq:parametrisation} from table~\ref{tab:characteristics} (or table~\ref{tab:characteristicsTrajectory}) within the quoted uncertainties to better than $0.03\%$ at quasi-unitarity ($0.4\%$ for the $3$rd excitation). At zero binding, the tension is about $0.6\%$ but still within uncertainties ($1\%$ for the ground state). In addition, the table compares to the direct determination of the slope $\calR^\jth$ with respect to $x_\kappa$ at quasi-unitarity, eq.~\eqref{eq:trajectoryToQU}, translated to the slope $(\rho^\jth)^\prime(\theta)$ via
\begin{equation}
    \frac{\dd\calR^\jth(x_\kappa)}{\dd x_\kappa}=
    \frac{(\rho^\jth)^\prime(\theta)\;\cos[\theta+\frac{\pi}{4}]+\rho(\theta)\;\sin[\theta+\frac{\pi}{4}]}
    {(\rho^\jth)^\prime(\theta)\;\cos[\theta+\frac{\pi}{4}]-\rho(\theta)\;\sin[\theta+\frac{\pi}{4}]}\;\;,
\end{equation}
at $\theta_\mathrm{QU}=\frac{\pi}{4}$ with $\rho_\mathrm{QU}=\kappaQU^\jth/(\sqrt{2}\kappathr^\jth)$ from table~\ref{tab:characteristicsTrajectory}. All results are in reasonable agreement.

All this makes us confident that the truncated parameters of table~\ref{tab:parametrisation} provide an adequate representation of the best fit. Accounting for all information, including the confidence and prediction uncertainty intervals of the fit as well as the comparison to the other extractions, an uncertainty of about $0.5\%$ overall seems reasonable, and somewhat more towards the zero-binding endpoints. This should in any case suffice to address physical systems, where experimental uncertainties and  higher-than-leading order corrections enter as well.

\subsection{Comparison to Short-Range EFT}
\label{sec:compare-Efimov}

The universality of Resummed-Range EFT as $j\to\infty$ can be understood from the discussion in sect.~\ref{sec:Efimov}. Still, it is instructive to consider in more detail the transition~\eqref{eq:2BpropEfimov} in the \twoB propagator from Resummed-Range to Short-Range EFT. Since $P,Q,\mu$ are dimensionless and $\mu^2=-\kappa_3^2<0$ for a bound \threeB state, one can differentiate two regions in the propagator of the integral equation~\eqref{eq:kernel}. For $3Q^2+4\kappa_3^2\gtrsim16$, the effective-range term $(3Q^2+4\kappa_3^2)/4$ cannot be neglected and may indeed dominate over the Short-Range EFT term of the propagator, $\sqrt{3Q^2+4\kappa_3^2}$. The integral converges quickly. The scale of the loop momentum is set by the binding momentum, $Q\simle\kappa_3$. As a consequence, we expect that bound states can be supported at $\kappa_3\lesssim 2$. Interestingly, this is not far from our numerical finding for the ground state, $\kappa_3^{(0)}\lesssim2.1$. In contrast, for $3Q^2+4\kappa_3^2\ll16$, the effective-range term becomes a correction while the propagator reduces to
\begin{equation}
  \label{eq:2Bproplimitbs}
  \frac{1}{\xi+(3Q^2+4\kappa_3^2)/4+\sqrt{3Q^2+4\kappa_3^2}}\to
  \frac{1}{\xi+\sqrt{3Q^2+4\kappa_3^2}}\to\frac{1}{\sqrt{3}\;Q}\;\;,
\end{equation}
and hence to the propagator of Short-Range EFT, for a shallow bound state when $|\xi|,\kappa_3\ll1$. Therefore, bound states are to be expected close to quasi-unitarity in Resummed-Range EFT, and the previous subsections confirm that with a tower of states near $\xi\approx0$. In general, this sequence of r\'egimes implies that range effects should be noticeable for lower-lying states in the spectrum. On top of that, we might expect the largest effect for the largest binding momenta, namely where the $3Q^2+4\kappa_3^2$ term should be largest. As we saw before, that is indeed near the point where the state disappears at the \twoB threshold.  

At a given $\xi$, $\kappa_2^+(\xi)$ remains fixed while the binding momenta of high excitations $j\to\infty$ approach zero and are therefore much smaller than the inverse effective range. We thus expect an expansion in powers of $|r_0|$,
\begin{equation}
  \label{eq:RRvsSR}
  \kappa_3^{(j)}(\xi(\theta)) = \kappa_\mathrm{Ef}^{(j)}(\theta) 
  \left[1+ b_{1}^{(j)}(\theta)\; 
  \kappa_\mathrm{Ef}^{(j)}(\theta) 
  + b_{2}^{(j)}(\theta)\; 
  (\kappa_\mathrm{Ef}^{(j)}(\theta))^2 +\ldots \right] \;\;,
\end{equation}
where $\kappa_\mathrm{Ef}^{(j)}(\theta)$ are the binding momenta for the Efimov tower at finite scattering length but vanishing effective range, but in units of $\absrm$, and $b_i^{(j)}(\theta)$ are functions which are universal for systems with large, negative effective range. Discrete Scale Invariance translates into equal spacing for the Efimov tower, keeping the angle $\theta$ unaffected but $\rho\to\e^{-\pi/s_0}\rho$. Therefore, it is sensible to use $\theta$ as independent variable in eq.~\eqref{eq:RRvsSR}. Thus, the ratio between adjacent states of the various properties of the special points (threshold, quasi-unitarity, and zero  binding) should approach the Efimov ratio $e^{\pi/s_0}$ as we go up the bound spectrum; see eqs.~\eqref{eq:Efimovratio}\ to~\eqref{eq:Efimovratioprime2}. 
The proximity of our tower to Efimov's can be seen in the lower rows of table~\ref{tab:characteristics}. Only the features of the ground state deviate by more than $10\%$ of the Efimov ratio, and only at threshold and zero binding. The deviations at these extreme points is ${\cal O}(\kappa_2^-/\kappa_2^+)$, as naively expected when  $\kappa_3^\jth=\kappa_2^-$ and $\kappa_3^\jth=0$, respectively. At quasi-unitarity, on the other hand, deviations seem smaller, ${\cal O}((\kappa_3^\jth/\kappa_2^+)^2)$. 

\begin{table}[!b]
  \centering
  \begin{footnotesize}
  \begin{tabular}{|l||l:l|l|l|}
    \hline
    \rule[-1.5ex]{0ex}{5.5ex}& \multicolumn{2}{c|}{ratio threshold to zero
                               binding} &
                                         \multicolumn{1}{c|}{\pbox{\linewidth}{ratio quasi-unitarity\\[-0.5ex]to zero binding}}&
                                         \multicolumn{1}{c|}{\pbox{\linewidth}{ratio threshold\\[-0.5ex]to quasi-unitarity}} \\
    \rule[-3ex]{0ex}{7.5ex}state& $\dis\frac{\xithr}{\xiZB}$& $\dis\frac{\kappathr}{\kappatwoZB}$& $\dis\frac{\kappaQU}{\kappatwoZB}$ &$\dis\frac{\kappathr}{\kappaQU}$\\\hline\hline
    ground        & $-23.783(1) $ & $-10.3692(3) $ & $-1.1526(1)$& $8.9963(1) $ \\\hline
    $1$st excited & $-19.185(3) $ & $-17.993(1) $ & $-1.482(1)$& $12.1413(1) $ \\\hline
    $2$nd excited & $-21.04(3) $  &  $-20.97(1) $ & $-1.506(3)$ & $13.921(3) $ \\\hline
    $3$rd excited & $-21.22(5) $   &   $-21.2(3) $ & $-1.51(3)$  & $14.05(2) $ \\\hline\hline
    $j\to\infty$  & \multicolumn{2}{l|}{\hspace*{8.3ex}$-21.3(2)$} & $-1.51(4)$ &$14.1(3)$ \\\hline\hline
    \rule[-2ex]{0ex}{5.5ex}Efimov $j\to\infty$& \multicolumn{2}{l|}{\hspace*{8.3ex}\pbox{\linewidth}{$-21.3048\dots$~\cite{Gogolin:2008NN} with~\cite{Braaten:2004rn}\\[-0.5ex]$-21.306(1)$\hspace{2.5ex}\cite{Deltuva:2012zy} with~\cite{Braaten:2004rn}}}& \pbox{\linewidth}{$-1.50763$~~~~~\cite{Gogolin:2008NN}\\[-0.5ex]
    $-1.5077(1)$\hspace{2.1ex}\cite{Deltuva:2012zy}} &\pbox{\linewidth}{$14.131\dots$\\[-0.5ex]$=(0.0707645\dots)^{-1}$}~\cite{Braaten:2004rn} \\\hline
  \end{tabular}
  \end{footnotesize}
  \caption{Characteristics of the trajectories from threshold to zero \threeB binding of the ground state and first three excitations, including their extrapolated values for the Efimov-like  states. Our ratios between quasi-unitarity and zero-binding are inferred from the other ratios. Uncertainties combine extrapolation and numerics. The last row lists the ratios in leading-order Short-Range EFT, with references.} 
  \label{tab:characteristicsTrajectory} 
  \end{table}

In addition, the universal trajectory $\calR^{(j\to\infty)}$ should become identical to the one of Short-Range EFT, whose characteristic values are well-explored~\cite{Gogolin:2008NN,Braaten:2004rn,Deltuva:2012zy}.  When these quantities are rescaled by the threshold parameters, universal numbers also emerge within each trajectory of the highly  excited levels. Indeed, there is agreement of our values at the $1\%$ level already for the second excitation, and excellent agreement  for the extrapolations $j\to\infty$. Table~\ref{tab:characteristicsTrajectory} shows that ratios between threshold and zero binding, $\kappathr^{(j)}/\kappatwoZB$ ($\xithr^{(j)}/\xiZB$), approach $\simeq -21.3$ as $j\to\infty$, compared with the literature value $\simeq  -21.305$~\cite{Braaten:2004rn,Gogolin:2008NN,Deltuva:2012zy}. Similarly, the ratios $\kappathr^{(j)}/\kappaQU^{(j)}$ between threshold and quasi- unitarity approach $\simeq 14.1$, while the literature value of Short-Range EFT is $\simeq 14.131$~\cite{Braaten:2004rn}. The two literature  values~\cite{Gogolin:2008NN,Deltuva:2012zy} for the ratio between quasi-unitarity and zero binding are identical within the error bar quoted in ref.~\cite{Deltuva:2012zy}, and our result agrees with both. This  confirms yet again that the EFT with non-perturbative effective-range corrections includes Short-Range EFT for sufficiently high excitations. Thus, we recover Discrete Scale Invariance far up in the bound spectrum. 

However, there is an important difference to Resummed-Range EFT. In Short-Range EFT, the position of the geometric tower of \threeB states is fixed by the dimensionful scale parameter $\LambdaEfimov$ of the \threeB interaction $H(\Lambda)$. In contrast, here the position of the tower is a parameter-free prediction once the \twoB pole positions are determined from, say, \twoB scattering. In other words, Resummed-Range EFT can be seen as an underlying theory to Short-Range EFT, \emph{predicting} $\LambdaEfimov$ in units of $\absrm$. The reason is, of course, that Short-Range EFT has no remaining natural scale once the \twoB binding is zero, while the second pole at $\kappa_2^+=-2$ still provides one in Resummed-Range EFT.

To illustrate this prediction, we focus on $\xi=0$. The binding momenta at quasi-unitarity are a particular case of eq.~\eqref{eq:RRvsSR} when $\theta=\pi/4$. The proximity of the ratio of binding momenta at quasi-unitarity and unitarity in table \ref{tab:characteristics} suggests the expansion at $\theta=\pi/4$ converges even for the ground state. We therefore computed $H(\Lambda)$ in Short-Range EFT at $\gamma=0$ such that the binding momenta $\kappa_3^\jth$ of the ground state or any of its first three excitations is identical to the results of Resummed-Range EFT at $\xi=0$ which are listed in table~\ref{tab:characteristics}. We employed $156$ values of $\Lambda\absr$ between $32$ and $16,384$, equidistant in $\ln\Lambda$. From these, we determined $\LambdaEfimov \absr$ using eq.~\eqref{eq:Hrunning}, setting it close to $1$ considering that it is by eq.~\eqref{eq:Hrunning} only unique only up to  multiples of $\e^{\pi/s_0}$. First, we performed a three-parameter fit to $\LambdaEfimov$, the amplitude $A$, and the frequency $s_0$. Since we found that $s_0=1.00623(1)$ is statistically identical to the predicted value of eq.~\eqref{eq:s0}, $s_0=1.006237\ldots$, we then used the predicted value of $s_0$ to eliminate one parameter. The results are reported in table~\ref{tab:Efimov}. They are sufficiently stable against varying numerics (grid size, mapping, spacing) and cutoffs that we are confident in the extraction error estimate. All fits have an adjusted coefficient of determination $R^2$ which deviates from $1$ by less than $10^{-8}$. The amplitude $A=0.87866(1)$ is the same within error bars for all states, and agrees with the value of $A=0.879$ first  reported in ref.~\cite{Braaten:2011sz}. The renormalisation scale $\LambdaEfimov \absr$ decreases minimally from the ground state to the  first excitation, even less to the second one, and is stable for the third excitation and beyond. We see no substantial quality-of-fit  differences for different levels. That (in regularisation with a ``hard'' cutoff)
\begin{equation}
\label{eq:LambdaEfimovDetermined}
    \LambdaEfimov^{(j\ge2)}= 0.610206(1)\;\absrm
\end{equation}  
is near-identical for all levels can be traced to the small deviations from the Efimov ratio at quasi-unitarity discussed in the preceding subsection; see also table~\ref{tab:characteristics}. 
For this value of the three-body parameter,
\begin{equation}
\label{eq:LambdaEfimovConsequence}
    \kappa_\mathrm{Ef}^{(0)}\left(\frac{\pi}{4}\right)\simeq e^{3\pi/s_0}\; \kappa_3^{(3)}\left(\frac{\pi}{4}\right) = 0.233814(6)\;\;,
\end{equation}  
from table~\ref{tab:characteristics}, which when multiplied by $e^{\pi/s_0}/2$ gives $2.65313(7)$, agreeing with the value of $2.653(1)$ quoted in ref.~\cite{Gogolin:2008NN}.

\begin{table}[!b]
  \centering
  \begin{footnotesize}
  \begin{tabular}{|l||c|c|}
    \hline
    state & scale parameter $\LambdaEfimov\;[\absrm]$ & amplitude $A$\\\hline
    \hline
     ground \hf($j=0$)        &$0.614379(2)$ & $0.87866(2) $\\\hline
     $1$st excited  \hf($j=1$)&$0.610223(1)$ & $0.87866(1) $\\\hline
     $2$nd excited  \hf($j=2$)&$0.610206(1)$ & $0.87866(1) $\\\hline
     $3$rd excited  \hf($j=3$)&$0.610206(1)$ & $0.87866(1)$\\\hline
     \hline
     $j\to\infty$  & $0.610206(1)$ & $0.87866(1)$\\\hline
  \end{tabular}
  \end{footnotesize}
  \caption{Scale parameter $\LambdaEfimov$ in $\absrm$ and amplitude $A$ of the \threeB interaction $H(\Lambda)$ at $1/a=0$ in Short-Range EFT with ``hard'' cutoff regularisation, fixed to reproduce the results of Resummed-Range EFT at quasi-unitarity ($\xi=0$). Uncertainties combine fit and numerics.}
  \label{tab:Efimov}
\end{table}

It is interesting to compare the deviations we observe in the Efimov spectrum or, equivalently, in $\LambdaEfimov$ for a ``hard'' cutoff regularisation, to Short-Range EFT at subleading orders, where $r_0\ne0$. In a specific physical system, where other contributions beyond the effective range exist, $\Lambda_\star$ would be fixed by the energy of a particular \threeB level, for example the ground state. The \twoB interaction corresponding to the range is then included in perturbation theory --- for complete NLO and N$^2$LO calculations, see refs.~\cite{Ji:2011qg} and~\cite{Ji:2012nj}, respectively. The coefficients $b_1^{(j)}(\theta)$  in eq.~\eqref{eq:RRvsSR} are the matrix elements of these perturbations of LO states. In first-order perturbation, a correction to the \threeB interaction is necessary for renormalisation, and the determination of $b_1^{(j)}(\theta)$ requires an additional datum, various choices of which are described in ref.~\cite{Ji:2011qg}. However, Discrete Scale Invariance requires 
\begin{equation}
    b_1^{(j)}(\theta)=b_1(\theta)
\end{equation}
to be the \emph{same} for all levels, as the LO states are the same apart from the scaling factor. 
In other words, if $r_0$ were to scale --- that is, made a spurion field in the language of Particle Physics ---
as the inverse of a momentum, shuffling \threeB levels, then the normalised $\kappa_\mathrm{Ef}^{(j)}(\theta)$ would be invariant and $b_1^{(j)}(\theta)$ could not depend on the level $j$~\cite{vanKolck:2017jon}. 
If we determine the NLO \threeB interaction by requiring one of the levels at unitarity to remain unchanged, then 
\begin{equation}
    b_1\left(\frac{\pi}{4}\right)\stackrel{!}{=}0
\end{equation}
and all levels remain unchanged~\cite{Platter:2008cx}. In a strict comparison of just effective-range effects, we match Short-Range EFT at LO to Resummed-Range EFT at LO by assuming the third (or a higher) excitation agree, thus determining $\Lambda_\star$ as done above. That implies that the range corrections are quadratic in $r_0$ at quasi-unitarity, which explains why deviations from the Efimov spectrum are ${\cal O}((\kappa_3^\jth/\kappa_2^+)^2)$, as already found in sect.~\ref{sec:compare-Efimov}. Such quadratic corrections appear at N$^2$LO in Short-Range EFT, \ie~the effective-range interaction enters in second-order perturbation theory. Renormalisation requires a new, two-derivative \threeB interaction~\cite{Bedaque:1998km,Bedaque:2002yg,Ji:2012nj}, which breaks Discrete Scale Invariance and results in coefficients  $b_2^{(j)}(\theta)$ which are not necessarily the same for all states. The new \threeB interaction strength can in fact be determined by fitting a second \threeB level at (quasi-)unitarity. Not surprisingly, binding energies in Resummed-Range EFT become increasingly different from those at unitarity in Short-Range EFT as we move towards the ground state of the quasi-unitarity spectrum. Approximating $\kappaQU^{(j)}$ by the $\left(\kappa_\mathrm{Ef}^{(j)}(\pi/4)\right)^2$ term, we find from eq.~\eqref{eq:RRvsSR} and table~\ref{tab:characteristics} for example for the ground state $b_2^{(0)}(\pi/4)\approx 0.1$, which is small but not totally unnatural. With $b_2^{(j\ge 1)}(\pi/4)$ of similar magnitudes, the quadratic correction in eq.~\eqref{eq:RRvsSR} quickly disappears in the numerical noise. The argument above does not require $b_1(\theta\ne\pi/4)$ to vanish, but even so corrections to the Efimov tower quickly become very small as we go up the spectrum. 

Indeed, we checked directly that the trajectory of the $3$rd excitation is practically identical to the trajectory of an Efimov state whose \threeB interaction $H$~\eqref{eq:Hrunning} is determined to reproduce the same binding energy at (quasi-)unitarity, $\gamma=\xi=0$. The small differences of $<0.4\%$ we find are within numerical and extrapolation uncertainties, and are dwarfed by the thickness of the lines in the plots. Therefore, the blue lines labelled ``$3$rd excited ($j=3$)'' in figs.~\ref{fig:rescaled} to~\ref{fig:parametrisation} can safely be identified with the Efimov trajectory, as was already anticipated. In particular, the difference between rescaled trajectories in fig.~\ref{fig:binding-thr} is indeed a difference between each trajectory and the Efimov trajectory; and the parametrisation of the $3$rd excitation in table~\ref{tab:parametrisation} is for the Efimov trajectory, within the uncertainties induced by the strong correlations between parameters. 

\begin{figure}[t]
\begin{center}
  \includegraphics[width=0.6\linewidth]
  {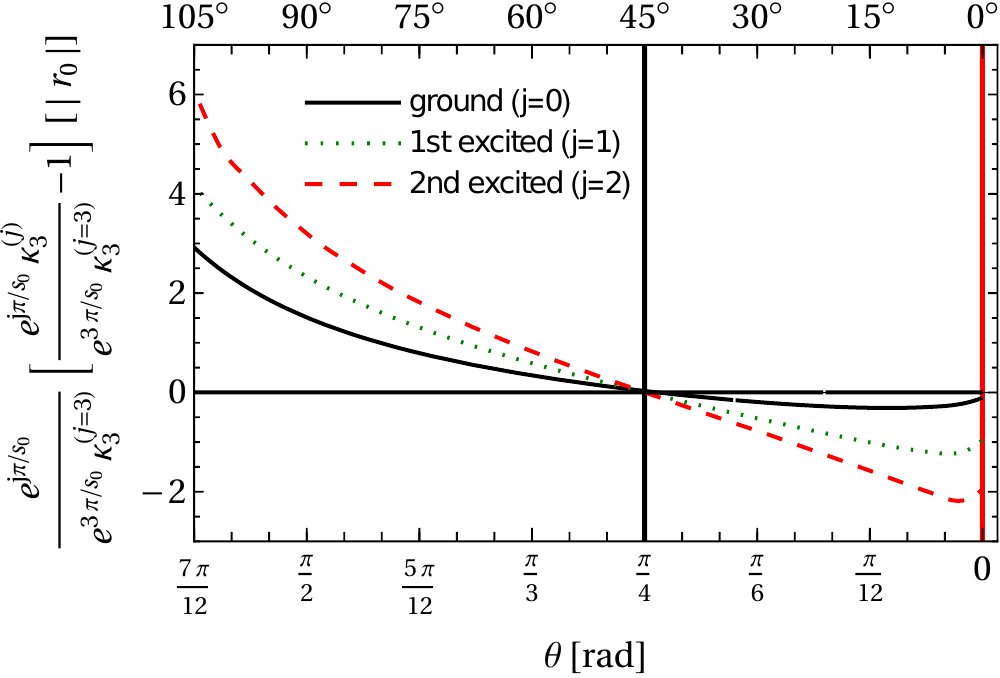}
    \caption{(Colour on-line) The ratio  $r^\jth(\theta)$, eq.~\eqref{eq:r(j)}, in the region where the numerics is considered stable for all states.}
  \label{fig:ratio-vs-theta}
\end{center}
\end{figure}

As we go down the spectrum, the deviations from Discrete Scale Invariance become more noticeable. One way to gauge the departure from the Efimov spectrum is through 
\begin{equation}
    r^\jth(\theta):= \frac{e^{j\pi/s_0}}{e^{3\pi/s_0} \kappa_3^{(3)}(\xi)} \left[\frac{e^{j\pi/s_0} \kappa_3^{(j)}(\xi)}{e^{3\pi/s_0}\kappa_3^{(3)}(\xi)}-1\right] 
    =  b_{1}(\theta) + b_{2}^{(j)}(\theta)\,\kappa_3^{(j)}(\xi(\theta))  +\ldots  \;\;,
    \label{eq:r(j)}
\end{equation}
which vanishes for $j=3$. This ratio measures the relative deviation between levels when all are rescaled to the ground state, in units of the binding momentum of the $3$rd state, rescaled to the $j$th state. It is plotted in fig.~\ref{fig:ratio-vs-theta} for the various states we consider. Towards zero binding, $\theta\to3\pi/4$, the binding momentum $\kappa_3^\jth(\kappa_2^{-\,\jth}\searrow\kappatwoZB[\jth])\to0$ of each state becomes increasingly small. That makes its subtraction from the (rescaled) $3$rd excitation in eq.~\eqref{eq:r(j)}
numerically increasingly unreliable. Therefore and in view of future studies of the details of $r^\jth(\theta)$~\cite{future}, we refrain from showing results for $\theta>7\pi/12=105^\circ$.

Following eq.~\eqref{eq:RRvsSR}, the linear correction in the effective range is approximately given by this ratio for the second excitation,
\begin{equation}
    b_1(\theta) \simeq r^{(2)}(\theta)  \;\;.
    \label{eq:r(2)}
\end{equation}
It decreases near-monotonically from zero binding to threshold, passing as expected through 0 at $\theta=\pi/4$. The first excitation has sizeable corrections from the quadratic term,
\begin{equation}
    r^{(1)}(\theta) - b_1(\theta) \simeq b_2^{(1)}(\theta) \; e^{2\pi/s_0} \; \kappa_3^{(3)}(\xi(\theta)) \;\;,
\end{equation}
where $e^{2\pi/s_0} \; \kappa_3^{(3)}(\xi)$ reaches 0.15 at threshold, using the numbers from table \ref{tab:characteristics}. As can be seen from fig.~\ref{fig:ratio-vs-theta}, the quadratic coefficient $b_2^{(1)}(\theta)$ changes sign at $\theta\approx \pi/4$ and always opposes the linear correction. While the expansion~\eqref{eq:RRvsSR} is expected to apply for the excitations, ground-state corrections can be large,
\begin{equation}
    r^{(0)}(\theta) - b_1(\theta) = b_2^{(0)}(\theta) \; e^{3\pi/s_0} \; \kappa_3^{(3)}(\xi(\theta)) + \ldots \;\;,
\end{equation}
since $e^{3\pi/s_0} \; \kappa_3^{(3)}(\xi<0)\simge 0.23$. In fact, we see from fig.~\ref{fig:allbindings} that this expansion should not converge once $\kappa_2^-\gtrsim\sqrt{2}-1\approx0.4$ ($\xi\simle -1$), where $\kappa_3^{(0)}\simge 1$. The effects of terms quadratic and higher in $r_0$ have the same qualitative behaviour as in the first excitation, but are larger.

Since $b_1(\theta)$ grows in magnitude away from $\theta=\pi/4$, we expect the largest corrections at the two other points of particular interest, namely zero-binding and threshold. This is indeed revealed by the larger deviations between Resummed-Range and Short-Range EFTs there. These deviations are, as one would expect, in the opposite direction to those found for (small) $r_0>0$ in Short-Range EFT~\cite{Platter:2008cx}. The deviations at these end points can be connected through universal numbers. Equation~\eqref{eq:RRvsSR} can be applied directly at threshold ($\theta=0$), where $\kappa_3^{(j)}=\kappa_\mathrm{thr}^{(j)}$, by setting  $\kappa_\mathrm{Ef}^{(j)}(0) = \gamma_\mathrm{thr}^{(j)}$ to the \twoB binding momentum where the $j$th \threeB excitation becomes stable in LO Short-Range EFT. At zero binding we can introduce a similar expansion,
\begin{equation}
  \label{eq:ZB}
  \kappa_{2\mathrm{ZB}}^{-(j)} = \gamma_\mathrm{ZB}^{(j)}
  \left[1+ b_{1}\left(\frac{3\pi}{4}\right)\; \gamma_\mathrm{ZB}^{(j)}
  +\ldots \right] \;\;,
\end{equation}
where $\gamma_\mathrm{ZB}^{(j)}$ is the (negative) \twoB binding momentum at which the $j$th \threeB excitation disappears in LO Short-Range EFT. The shifts at the two end points relative to the unitarity limit are then 
\begin{equation}
  \label{eq:ZB-Thr}
  \frac{\kappa_{2\mathrm{ZB}}^{-(j)}}{\gamma_\mathrm{ZB}^{(j)}} - 1
  = \frac{b_{1}(\frac{3\pi}{4})}{b_{1}(0)}\; 
  \frac{\gamma_\mathrm{ZB}^{(j)}}{\gamma_\mathrm{thr}^{(j)}}
  \left(\frac{\kappa_\mathrm{thr}^{(j)}}{\gamma_\mathrm{thr}^{(j)}}-1\right)
  +\ldots \;\;.
\end{equation}
A similar relation was obtained in ref.~\cite{Ji:2011qg}.

After this first exploration of the bound spectrum, of its discrete scaling properties, of the trajectories of each state as $\xi=2r_0/a$ (or, equivalently, $\kappa_2^-$) is varied, and of the emergence of the Efimov tower in Resummed-Range EFT, we now turn to scattering.

\section{Some Scattering Results}
\label{sec:scattering}

The differences in bound states compared to Short-Range EFT prompts us to look at the scattering  of one boson on the bound state of the other two which exists for $\kappa_2^->0$ ($\xi<0$). A detailed exploration of scattering is left for a future publication~\cite{futureScattering}; here, we highlight some of its rich structure and concentrate again on the transition to Short-Range EFT.

The scattering equation eq.~\eqref{eq:3Bamprescaled} is solved using the Hetherington-Schick method on a contour in the fourth quadrant of  the complex $P,Q$ plane, inside the wedge $0\ge\Im[P,Q]\ge-2\kappa_2^-\;\Re[P,Q]/K$ where the amplitude is known to be  analytic~\cite{HetheringtonSchick, AaronAmado, CahillSloan, Brayshaw, SchmidZiegelmann}. Its advantage is that it readily applies to  momenta $K$ above the breakup threshold 
\begin{equation} 
    \label{eq:breakup}
    K_\mathrm{thr}:=\sqrt{\frac{4}{3}}\;\kappa_2^-
\end{equation} 
where phase shifts and $K\cot\delta$ become complex. The underlying code has already been used in refs.~\cite{Bedaque:1999vb, Gabbiani:1999yv, Bedaque:2002yg, Griesshammer:2004pe, Griesshammer:2011md}. We use up to $n=1300$ points with a quasi-exponential mapping which weighs more both small $P,Q$ and the region around $K$. The results are stable against using different  grids, mappings and contours, although arcs of circles starting at $(P,Q)=0$ and ending at cutoff $\Lambda$ appear to provide the most  stability. We chose to show results for one representative ``high'' cutoff of $\Lambda=45.2548\dots\absrm$; results with half this value are nearly indistinguishable even at the high end of the $K\le11.3137\dots$ we report on, that is, up to a quarter of the cutoff. More  details at low $K$ are available from plots of $K\cot\delta$ since phase shifts necessarily go to zero (modulo $\pi$) for $K\to0$. In particular,  one recovers the effective-range expansion,
\begin{equation}
    \label{eq:effrange3body}
    K\cot\delta(K\to0)=-\frac{1}{a_3}+\frac{r_3}{2}\;K^2+\calO(K^4)\;\;,
\end{equation}
where $a_3$ and $r_3$ are the scattering length and effective range for $\B(\B\B)$ scattering, measured yet again in units $\absr$ of the effective range of the \twoB system.

To explore similarities and differences with Short-Range EFT, it is instructive to consider more quantitatively the $\kappa_2^-\to0$ limit of the propagator of the integral equation~\eqref{eq:kernel}, as we did for bound states but now using $\mu^2=3K^2/4-(\kappa_2^-)^2$. The integral equation is dominated by momenta $Q$ for which its \twoB propagator (remember the relation $\xi=-\kappa_2^-(\kappa_2^-+2)$) 
\begin{equation}
    \label{eq:propininteq}
    \frac{1}{\xi +\frac{3}{4}(Q^2-K^2)+(\kappa_2^-)^2+2\sqrt{\frac{3}{4}(Q^2-K^2)
    +(\kappa_2^-)^2}}
\end{equation}
is large, namely only for $Q\sim K$. It is this integrable singularity of the \threeB amplitude at $Q=K$ on the real axis which makes the complex contour of the Hetherington-Schick method so attractive. Similarly to the bound-state case, we expect the range contribution to be relatively small in the limit $\kappa_2^-\to0$ when $3(Q^2-K^2)\simle 16$, which translates into $Q\sim K\simle 2\sim |\kappa_2^{+}|$. As we obtained before, the deeper bound state sets the momentum scale for the transition from Resummed-Range to Short-Range EFT.

\begin{figure}[t]
\begin{center}
  \includegraphics[width=\linewidth]
  {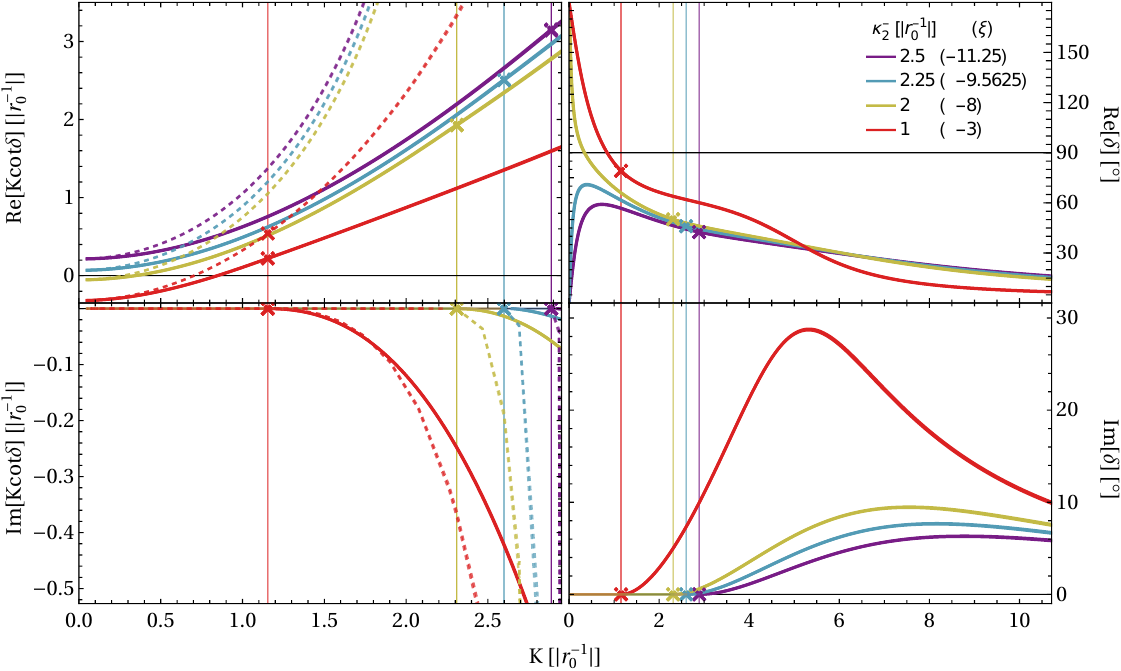}
    \caption{(Colour on-line) Scattering of one boson from a bound two-boson system with resummed effective range for parameter sets around the threshold at which the \threeB ground state emerges from the continuum ($\kappathr^{(0)}=2.11\ldots$, $\xithr^{(0)}=- 8.72\ldots$). Left: $K\cot\delta$ at low momenta $K$. Right: phase shifts for momenta $K$ up to $\Lambda/4$. Top: real parts. Bottom:  imaginary parts, nonzero only above the $\B(\B\B)\to\B\B\B$ disintegration threshold, whose position $K_\mathrm{thr}$ at a given $\kappa_2^-$ is denoted by colour-coded crosses and vertical lines. Solid lines: Resumed-Range EFT. Dashed: Short-Range EFT.}
\label{fig:AroundGroundState}
\end{center}
\end{figure}

Figure~\ref{fig:AroundGroundState} shows the results when the \twoB binding is around the threshold point, namely where the \threeB ground state emerges from the continuum ($\kappa_2^-\sim\kappathr^{(0)}\approx 2$).
In the sub-threshold region, $\kappa_2^->\kappathr$, the scattering length of the $\B(\B\B)$ system is negative, $K\cot\delta(K\to0)>0$, and $\delta(K \to0)\to 0$, as expected for a system without a \threeB bound state. Conversely, in the all-bound region close to the first  threshold, the $\B(\B\B)$ scattering length is positive, $K\cot\delta(K\to0)<0$, and $\delta(K\to0)\to \pi$, indicating the presence of a bound \threeB state. Both real and imaginary parts of $K\cot\delta$ have monotonic behaviours in this momentum region, but the results in Short-Range EFT diverge sharply from those of Resummed-Range EFT for $K\gtrsim 2$, as expected. 

\begin{figure}[!t]
\begin{center}
  \includegraphics[width=0.5\linewidth]
  {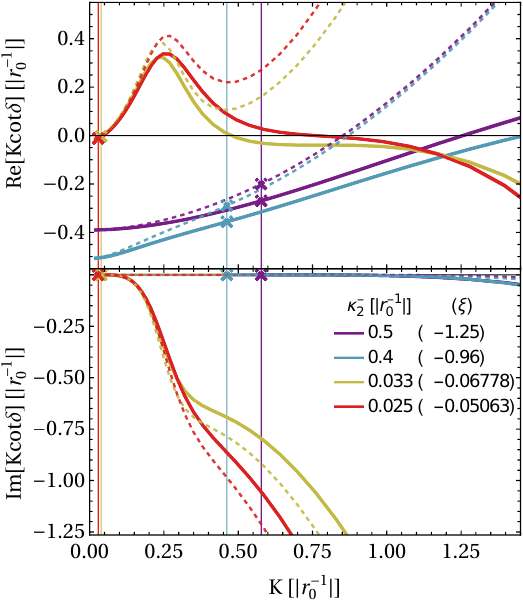}
    \caption{(Colour on-line) Real (top) and imaginary (bottom) part of $K\cot\delta$ at low momenta $K$ for small $\kappa_2^-(\xi)$ but  not close to thresholds $\kappathr^{(0)}\approx2.1$, $\kappathr^{(1)}\approx0.12$ or $\kappathr^{(2)}\approx0.006$. Colour coding, dashing, crosses and vertical lines as in fig.~\ref{fig:AroundGroundState}.}
\label{fig:GenericPointsLowXi}
\end{center}
\end{figure}

As $\kappa_2^-$ decreases, so does the break-up threshold, and non-monotonicity appears in $K\cot\delta$ at smaller momenta --- momenta well inside the region where Short-Range EFT is expected to hold. In fig.~\ref{fig:GenericPointsLowXi}, we address binding momenta $\kappa_2^-\in\{0.5;0.4\}$ in the all-bound region between the threshold for the bound ground state ($\kappathr^{(0)}\approx2.1$) and for the first excitation ($\kappathr^{(1)} \approx0.1$), as well as $\kappa_2^-\in\{\frac{1}{30};\frac{1}{40}\}$ between that for the first and second excitation ($\kappathr^{(2)}\approx\frac{1}{160}$). In both cases, discrepancies to Short-Range EFT become prominent at a somewhat lower scale $K\approx 1$. It is particularly remarkable that both EFTs capture the ``hump'' around $K\approx0.25$ for $\kappa_2^-\in\{\frac{1}{30};\frac{1}{40}\}$.

\begin{figure}[!p]
\begin{center}
  \includegraphics[width=\linewidth]
  {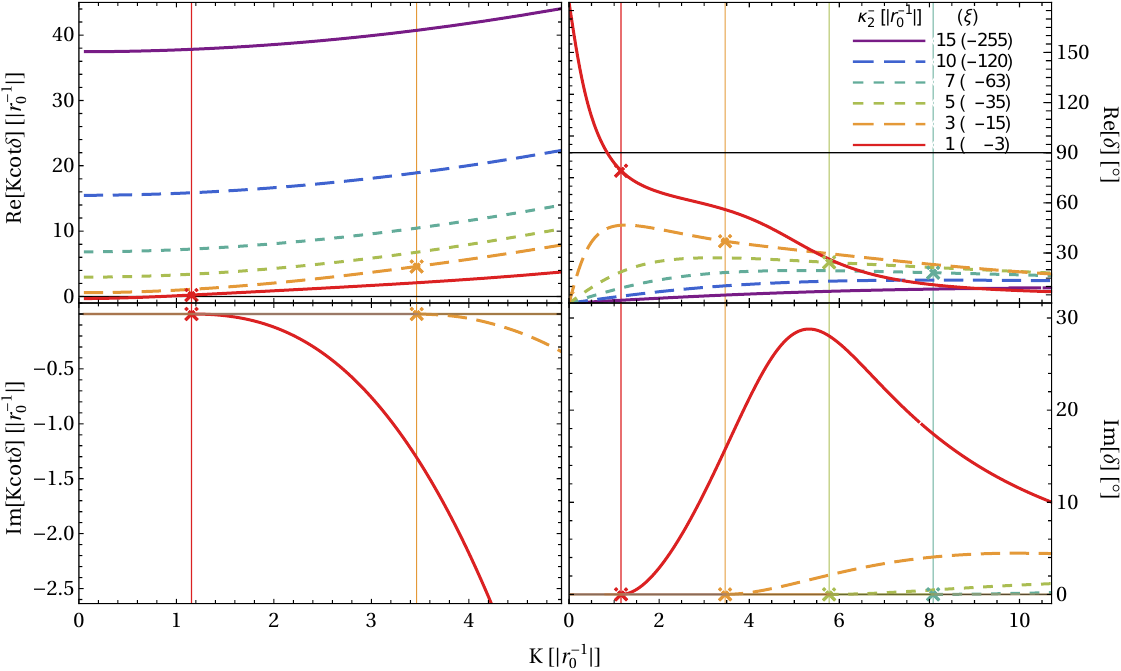}
\\[3ex]
  \includegraphics[width=\linewidth]
  {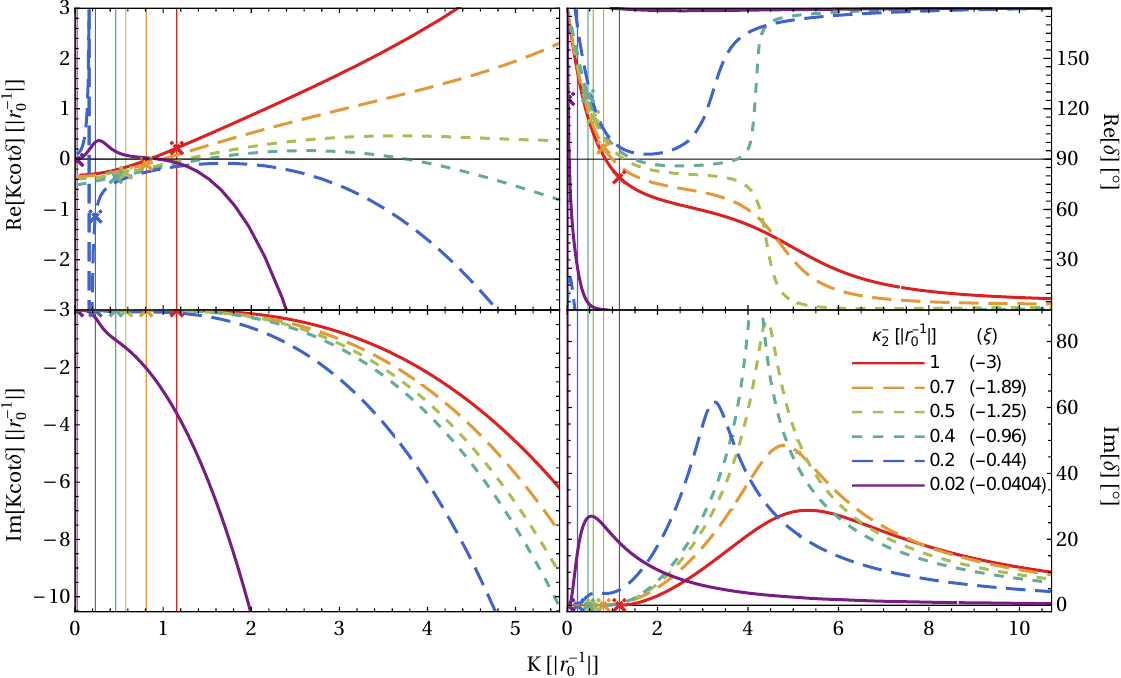}
\caption{(Colour on-line) Evolution of $K\cot\delta$ at low momenta (left) and of the phase shift $\delta$ up to $K=\Lambda/4 $ (right) for $\kappa_2^-\in[15;1]$ ($\xi\in[-255;-3])$ (top) and $\kappa_2^-\in[1;0.02]$ ($\xi\in[-3;-0.\overline{04}])$ (bottom). The red solid line  ($\kappa_2^-=1$, $\xi=-3$) is shown on both plots for orientation. Colour-coded vertical lines and crosses denote the respective breakup threshold.}
\label{fig:GenericPointsSpanningRange}
\end{center}
\end{figure}

Finally, fig.~\ref{fig:GenericPointsSpanningRange} scans phase shifts and low-momentum $K\cot\delta$ for generic values of $\kappa_2^-$ from well above $\kappathr^{(0)}$ for the ground state to before the emergence of the first excitation (top figure), and from there to just above the second excitation (lower figure). For large $\kappa_2^-$, phase shifts are small, $a_3<0$ is small in magnitude, and the amplitude could be treated with the Born approximation. That conforms with the estimate that for  $\kappa_2^-\gg K$, $Q^2-K^2$ can be treated around $\delta(K\to0)=0$ as perturbation in eq.~\eqref{eq:propininteq}. As the first  threshold $\kappathr^{(0)}\approx2.1$ is crossed, $K\cot\delta(K\to0)$ and the scattering length change sign, and $\delta(K\to0)\to \pi$, as discussed above. As the breakup threshold lowers, imaginary parts become much bigger. 

As $\kappa_2^-$ approaches from above the threshold $\kappa_\mathrm{thr}^{(1)}\simeq 0.125$ for the emergence of the first \threeB excitation, interesting structures appear in the real and imaginary parts of the phase shifts. They are all stable against variation in the number of points, complex contour, exponential mapping and cutoff variation. The ``shoulder'' in $\Re [\delta]$ for $\kappa_2^-=1$ around $K\approx3$ becomes a prominent plateau as $\kappa_2^-\to0.5$ decreases even further. Between $0.5$ and $0.4$, namely at about $0.43$ which is close to the logarithmic mean $\kappa_2^-\approx0.46$ of $\kappathr^{(0)}$ and  $\kappathr^{(1)}$, the plateau turns into a valley: For smaller $\kappa_2^-$, $\Re [\delta]$ does not approach $0$ as $K\to\infty$, but $\pi$. Remarkably, this elongated flat structure from $K\approx1$ up to about $4$ has $\Re [\delta]\simeq 80^\circ$. It is thus very close to saturating the unitarity bound ($\delta=90^\circ$), albeit all of the subsystems have nonzero binding energies. The plateau (valley) in $\Re [\delta]$ is followed by a sharp decrease (increase), which is correlated with a sharp peak in $\Im [\delta]$. With decreasing $\kappa_2^-$, it wanders to larger $K$s and its height increases, correlated with an increasing maximum slope. Large $\Im [\delta]$ corresponds to a small elasticity parameter $\eta$, thus we are witnessing a trough in $\eta$. This is a marker for a resonance in a multi-channel system, as an $S$-matrix pole at complex energy $E_\mathrm{R}-i\Gamma_\mathrm{R}/2$ leads to a minimum of $\eta$ at centre-of-mass energy $E_R$ with width determined by $\Gamma_\mathrm{R}$~\cite{GauthierKamal}. However, since the structure cannot adequately be captured by a Breit-Wigner form with constant background as suggested in ref.~\cite{Ceci:2013zta} and since we do at present not locate a complex pole directly, we leave it for future work to confirm the presence of resonances, map their trajectory in the complex $K$ plane, and relate them more directly to the emergence of excitations~\cite{futureScattering}. To summarise the issue, we have thus far been unable to find an intuitive explanation for these three features of the phase shift in a range of $\xi$ values: near-constant in a rather large momentum region; there, tantalising close to the unitarity bound of $90^\circ$; and there paired with a rather large inelasticity. Therefore, we here simply report the phenomenon.

Around $\kappa_2^-\approx0.2$ close to the emergence of the first excitation from the continuum, the scattering  length changes sign so that again $\delta(K\to0)\to0$ for even smaller $\kappa_2^-$. At the same time, the phase shifts drop dramatically from $\delta(0)=\pi$
to $0$ at very small $K$. That implies that the effective range of $\B(\B\B)$  scattering  becomes very large, as also seen from the large curvature in $K\cot\delta(K\to0)$. In fact, $\Re[K\cot\delta]$ blows up just before threshold, indicating a zero of the scattering amplitude. Note that the phase shift approaches the asymptotic free-particle limit $\delta\to0$ or $\pi$ much more rapidly for large $K$ as $\kappathr\searrow0$ is decreased. Indeed, for  $Q^2,K^2\gg Q,K\gtrsim\kappa_2^-$, eq.~\eqref{eq:propininteq} reduces at leading order to $4/[3(Q^2-K^2)]$  in momentum space. This suggests that phase shifts rapidly become perturbative as $K\to\infty$.

When plotted as a function of $\xi$~\cite{PhysRevLett.93.143201}, the $\B(\B\B)$ scattering length $a_3$, eq.~\eqref{eq:effrange3body}, shows the standard behavior as it varies from $-\infty$ to $+\infty$ between successive values of $\xi_\mathrm{thr}^{(j)}$ (\ie~of $\kappa_2^-$). While the \twoB effective range is by definition always negative to provide a well-defined EFT, the effective range $r_3$ is always positive. We did not find any $\xi$ ($\kappa_2^-$) for which it is negative.


\section{Summary and Conclusions}
\label{sec:conclusions}

We studied the three-boson bound states and scattering in an Effective Field Theory with two shallow two-body $S$-wave poles~\cite{Habashi:2020qgw,Habashi:2020ofb,vanKolck:2022lqz} at leading order. At this order, properties of few-body systems are determined by two parameters, say the two-body effective range $r_0$ and the dimensionless ratio $\xi=2r_0/a$, where $a$ is the \twoB scattering length. The emerging Resummed-Range EFT is self-consistent and does not need a three-body interaction at leading order to be renormalisable. 
At LO, it reproduces results that were already found for ``narrow Feshbach resonances''~\cite{PhysRevLett.93.143201} in Atomic Physics. 

In this EFT, the effective range provides a lower bound for the spectrum, and three-body bound states are found only in the range $\xithr\simeq -8.726 < \xi < \xiZB\simeq 0.367$ for the ground state ($2.119>\kappa_2^->-0.204$).  We see no indications for bound three-body systems with values $\xi$ ($\kappa_2^-$) outside that narrow window. In particular, this is still far away from the point $\xi=1$ ($\kappa_2^-=-1$) where the two virtual two-body states collide to form a resonance.

At first glance, the spectrum appears quite close to a truncated Efimov spectrum, but noticeable differences exist for the low-lying states; see \eg~table~\ref{tab:characteristics} and fig.~\ref{fig:binding-thr}. In particular, the ground state's threshold point (where \threeB and \twoB states have the same binding energy) is with $\kappathr^{(0)}\approx2.12\;\absrm$ about $30\%$ smaller than in the Efimov case, where it is at about $3.31\;\absrm$. The difference is still about $15\%$ for the first excitation. Conversely, the zero-binding point of the ground state is about $30\%$ further away than that of an Efimov state, and that of the first excitation, about $10\%$. Thus, the ground state and lower excitations become unstable against two-body decay considerably earlier than in the Efimov version, while they survive considerably longer as bound states when the Efimov state is already virtual. Overall, the parametrisation in sect.~\ref{sec:parametrising} shows that trajectories of the ground state and first few excitations are considerably shorter in $\kappa_2^-$ than that of a corresponding Efimov state. Furthermore, the three-body ground-state momentum at quasi-unitarity (vanishing shallowest two-body binding momentum) is only $\simeq 0.2\;\absrm$. It becomes comparable to or larger than $\absrm$ only for $\xi\simle -3$ ($\kappa_2^-\gtrsim \absrm$). Excitations, when they exist, all have momenta $\simle 0.1\;\absrm$. 

Consequently, the ground state and first excitation are close to threshold not well described with Short-Range EFT~\cite{Hammer:2019poc} (with a single low-energy two-body state). The higher spectrum is more amenable, if one is only interested in accuracies on the level of a few per-cent. Even for the excited spectrum, there is a remarkable difference between Resummed-Range and Short-Range EFT. In the latter, the relative position of the levels is fixed by Discrete Scale Invariance, while the absolute position requires a three-body parameter since no two-body scale is left. In contradistinction, in Resummed-Range EFT, this parameter is determined by the effective range, or, equivalently, by the binding momentum $\kappa_2^+(\xi=0)=-2\;\absrm$ of the second, ``deep'' bound two-body state even when $\kappa_2^-(\xi=0)=0$. This natural two-body scale remains, in contrast to unitarity in Short-Range EFT which has no two-body scale at all.  This is the most striking difference between many-body systems in the two versions. 

A main finding of our investigation is therefore: If one interprets Short-Range EFT as low-energy version of Resummed-Range EFT, then the absolute position of the Efimov tower is fixed, and the corresponding Efimov scale is determined as $\LambdaEfimov=0.610206(1)\;\absrm$ (in a renormalisation scheme with ``hard'' cutoff regularisation); see eq.~\eqref{eq:LambdaEfimovDetermined}. It will be quite interesting to put this prediction to the test in systems where Resummed-Range EFT applies.

Larger differences with Short-Range EFT appear in scattering at momenta comparable with $\absrm$. This leads to quite interesting structures in the phase shifts whose predictions could also be explored in physical systems. There are, for example,  plateaux for certain ranges of $\xi\approx0.5$ where the real part of the phase shift is near-independent of the scattering momentum $K$ and close to the unitarity limit $\delta\approx90^\circ$ albeit none of the subsystems are scale invariant. These plateaux are followed by sharp cliffs or walls, which are correlated with maxima of the imaginary part of the phase shifts --- a potential sign of a resonance. Remarkably, we did not find a parameter range for which the three-body effective range is negative, while the two-body one always must be to obey the Wigner bound. 

Indeed, we expect that the size of deviations of Resummed-Range EFT from the Short-Range EFT, and hence from Discrete Scale Invariance, grows larger in the spectrum of more-boson systems. Its upper part should still display Discrete Scale Invariance, for example the four-boson system should have two states associated with each Efimov state~\cite{Hammer:2006ct}. But that is not necessarily the case for the lower part, and in particular for the ground state. At unitarity in Short-Range EFT, the ground-state binding energy per particle grows nearly linearly with the number of bosons~\cite{vStecher:2010NN,Carlson:2017txq}, before tapering off according to liquid-drop-type saturation~\cite{Carlson:2017txq}. It is likely that ground states will have momenta comparable to or larger than $\absrm$. At unitarity, the four-boson ground state, for example, is already more bound than the three-boson ground state by a factor of about $4.6$~\cite{Hammer:2006ct}. At quasi-unitarity, the factor will be different but, if it is not very different, the four-boson ground-state binding momentum should still be close to $\absrm$. Moreover, the evolution of the binding energies per particle as function of particle number will now depend on $\xi$, in such a way that $\xi\to 0$ should approach the unitarity results of ref.~\cite{Carlson:2017txq}. Further work to clarify the effects of $r_0<0$ and large $|a|$ on the spectrum of more-boson systems would be of high interest.

Obvious further extensions include systems of three identical fermions; bosons with different masses; more than three particles; a more detailed study of scattering states including when none of the two-body systems is bound; and a study of the virtual and resonance states of the three-body system, especially for $\xi$ ($\kappa_2^-$) outside the range which allows for bound states (including two-body resonances). It is  of course desirable to confront our predictions with data of suitable systems with large but negative effective two-body effective range and large-in-magnitude scattering length in Atomic, Nuclear and Particle Physics. Before that, confidence in the self-consistency of the theory should be strengthened by the study of higher-than-leading-order effects. In particular, at NLO an interaction that accounts for the two-body shape parameter in first-order perturbation theory has been shown to improve the description of a toy model~\cite{Habashi:2020qgw,Habashi:2020ofb}. The renormalisation of this interaction in the three-body system might require a three-body force, which---being perturbative---cannot cause a collapse. These corrections should be small for low-energy observables. Work in this direction is in progress.


\section*{Acknowledgements}
We are particularly indebted to Lorenzo Contessi, Alireza Dehghani, Hans-Werner Hammer, Johannes Kirscher, Sebastian K\"onig, Manolo Pav\'on Valderrama, and Daniel R.~Phillips for discussions. 
Johann Haidenbauer, Christoph Hanhart, and Felix Werner provided context on physical systems with potentially large and negative effective range.
Instrumental for this research were the warm hospitality and financial support for HWG's stays at IJCLab Orsay, UvK's stay at George Washington University, and for both stays at the Kavli Institute for Theoretical Physics which is supported in part by the National Science Foundation under Grant No.~NSF PHY-1748958.
This material is based upon work supported in part by the U.S.~Department of Energy, Office of Science, Office of Nuclear Physics, under awards DE-SC0015393 (HWG) and DE-FG02-04ER41338 (UvK). 

\section*{Data Availability Statement}

All data underlying this work are available in full upon request from the authors.


\appendix
\setcounter{section}{0} 
\renewcommand\thesection{\Alph{section}}  
\setcounter{figure}{0}
\renewcommand\thefigure{\thesection.\arabic{figure}}  
\setcounter{equation}{0}
\renewcommand\theequation{\thesection.\arabic{equation}}  
\setcounter{table}{0}
\renewcommand\thetable{\thesection.\arabic{table}}

\section{Appendix: Trajectory Parametrisation of Sect.~\ref{sec:parametrising}}
\label{app:fit}
Our determination of the coefficients differs somewhat from Gattobigio \etal~\cite{Gattobigio:2019eqw}, who optimise the fit constrained by a  universal number of significant figures. We first perform an unconstrained fit for all coefficients $c_n$, $n=1,\dots,7$. The resulting $c_1$ to $c_4$ are then rounded to $5$ significant figures. With these coefficients fixed, we then re-fit the other ones, $c_5$ to $c_7$, and truncate that result at $6$ significant figures. Finally, we compare $\rho_\mathrm{param}$ of the final fit, with all coefficients truncated as prescribed, to that of the first fit, in which none of the coefficients were rounded beyond machine precision. The difference between first and final fit function is never bigger than $\pm1.5\cdot10^{-5}$ for each state.  The magnitude of the relative deviation between the two is $\lesssim\pm0.2\%$. At quasi-unitarity, truncated and free version are near-identical for all states. The estimated variance (root of the sum-squared of all residuals) is $\{0.00116;0.00039;0.00033;0.0015\}$, ordered from the ground state to the highest excitation. This measure worsen by $\lesssim0.02\%$ from the un-truncated to the truncated parametrisation ($0.17\%$ for the $2$nd excitation). The adjusted coefficient of determination $R^2$  worsens by $\lesssim0.06\%$ ($0.3\%$ for the $2$nd excitation), and the Bayesian Information Criterion (in the Gau\3ian approximation) by $\lesssim0.002\%$ ($0.016\%$ for the $2$nd excitation). In both the first and final parametrisations, the residuals are very similar for all states. Variants of this procedure, like different numbers of significant figures on different parameter combinations or iterative fits, lead to results of at best similar quality for the same overall number of significant figures --- and worse quality for smaller numbers.

\begin{figure}[!p]
\begin{center}
  \includegraphics[width=0.7\linewidth]
  {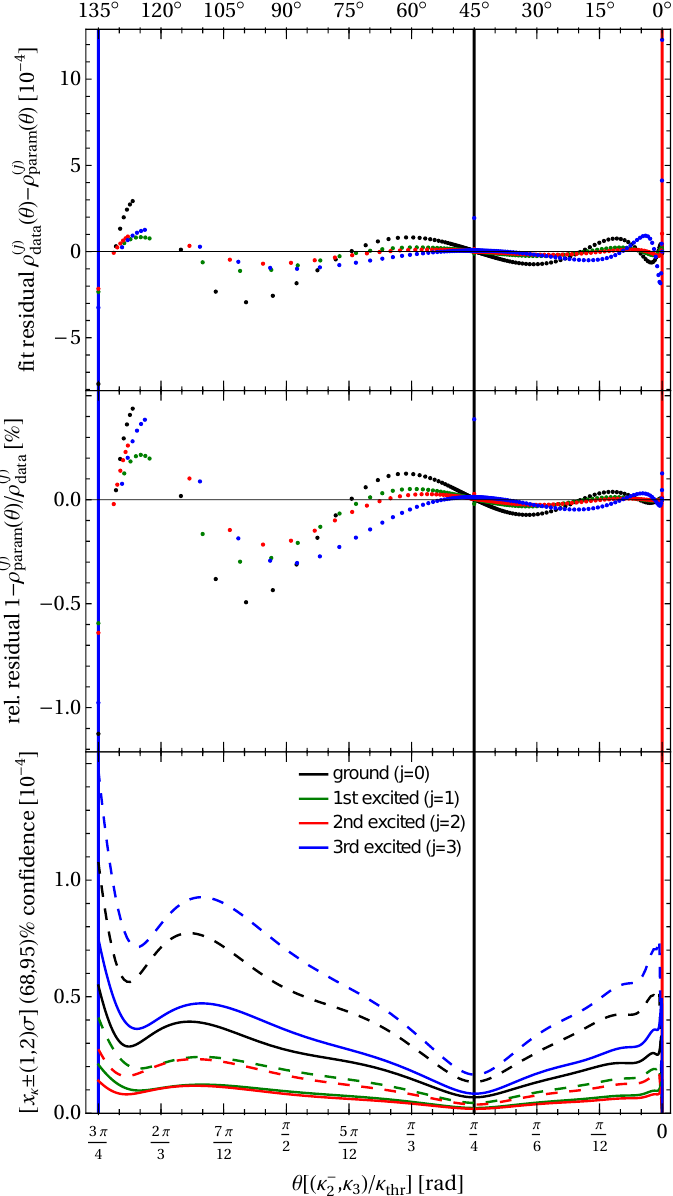}
    \caption{(Colour on-line) Quality-of-fit assessment for the trajectory of the ground state and first three excitations in polar form. The threshold is on the right; $\theta$ increases to the left. Top: residuals (rescaled by $10^4$). Centre: relative residuals. Bottom: function $w^\jth_\mathrm{conf}(\theta)$ (rescaled by $10^4$) parametrising the $68\%$ (solid lines) and  $95\%$ (dashed) confidence interval of the uncertainty around the mean, $\rho_\mathrm{param}^\jth(\theta)\pm w_\mathrm{conf}^\jth(\theta)$.}
\label{fig:parametrisationerror}
\end{center}
\end{figure}

We now discuss the accuracy of the fit result.
As fig.~\ref{fig:parametrisationerror} demonstrates, the residuals increase towards the threshold to $\le0.001$ in magnitude. While $\rho$ is big there, the relative residuals are with  $\pm0.1\%$ actually smaller than in the other regions. Otherwise, residuals are never bigger than $0.00005$ in magnitude. Relative residuals can reach $0.5\%$ in magnitude in the Borromean region around $105^\circ$. The $68\%$ confidence interval $\rho_\mathrm{param}^\jth(\theta)\pm  w_\mathrm{conf}^\jth$ of the uncertainty of the mean shows substantial dependence on $\theta$. It is minimal around quasi-unitarity and maximal at zero binding; see fig.~\ref{fig:parametrisationerror}. That reflects the high number of high-quality points around the former, and the notorious difficulty to  adequately map the approach $\kappa_3\to0$ for the latter. On top of that, it must by construction vanish at $\theta=0$, but the rapid fall bears yet again witness to the difficulty close to threshold to numerically uniquely disentangle a \threeB energy very close to the \twoB energy. Remember that this confidence interval captures only the variation in parameter estimates, while the prediction interval $w_\mathrm{pred}$  of table~\ref{tab:parametrisation} in addition accounts for the variation induced by new data and is thus considerably less $\theta$ dependent and, as expected, larger than $w_\mathrm{conf}$. 

Other reasonable parametrisations with the same number of parameters and similar rounding prescriptions show qualitatively similar behaviour but more pronounced fine-tuning and correlations. Clearly, the detailed numbers for the coefficients, for the uncertainty estimates and for the fit criteria depend significantly on the number and positions of the angles $\theta_i$ at which ``data'' for the fit are produced. For example, a fit with very many data close to threshold usually leads to distorted parametrisations which are poor at quasi-unitarity and zero binding. Overall, we believe that the distribution of points shown in the residuals of fig.~\ref{fig:parametrisationerror} is an adequate compromise for the present investigation.

\newpage
\end{document}